\documentclass[preprintnumbers,aps,11pt,nofootinbib,tightenlines,floatfix,notitlepage]{revtex4-1}
\usepackage[paperwidth=210mm,paperheight=297mm,centering,hmargin=1.95cm,vmargin=3.150cm]{geometry}

\usepackage{amsmath,amssymb}
\usepackage{graphicx}
\usepackage{bm}
\usepackage{comment}
\usepackage{color}
\usepackage{subfigure}
\usepackage{array}
\usepackage{multirow}
\usepackage{natbib}
\usepackage{url}
\usepackage[]{units}
\usepackage[T1]{fontenc}
\usepackage[latin9]{inputenc}
\usepackage{graphicx}
\usepackage{esint}
\usepackage{hyperref}
\usepackage{atbegshi}
\usepackage{lipsum}
\usepackage{braket}
\usepackage[normalem]{ulem}
\usepackage{soul}

\begin{document}

\dimen\footins=5\baselineskip\relax

\preprint{\vbox{
\hbox{MIT/CTP-4723}
}}

\title{Composite Vector Particles in External Electromagnetic Fields
}
\author{Zohreh Davoudi{\footnote{\tt davoudi@mit.edu}}
}
\affiliation{Center for Theoretical Physics, Massachusetts Institute of Technology, Cambridge, MA 02139, USA}

\author{William Detmold{\footnote{\tt wdetmold@mit.edu}}
}
\affiliation{Center for Theoretical Physics, Massachusetts Institute of Technology, Cambridge, MA 02139, USA}


\begin{abstract} 
Lattice quantum chromodynamics (QCD) studies of electromagnetic properties of hadrons and light nuclei, such as magnetic moments and polarizabilities, have proven successful with the use of background field methods. With an implementation of nonuniform background electromagnetic fields, properties such as charge radii and higher electromagnetic multipole moments (for states of higher spin) can be additionally obtained. This can be achieved by matching lattice QCD calculations to a corresponding low-energy effective theory that describes the static and quasi-static response of hadrons and nuclei to weak external fields. With particular interest in the case of vector mesons and spin-1 nuclei such as the deuteron, we present an effective field theory of spin-1 particles coupled to external electromagnetic fields. To constrain the charge radius and the electric quadrupole moment of the composite spin-1 field, the single-particle Green's functions in a linearly varying electric field in space are obtained within the effective theory, providing explicit expressions that can be used to match directly onto lattice QCD correlation functions. The viability of an extraction of the charge radius and the electric quadrupole moment of the deuteron from the upcoming lattice QCD calculations of this nucleus is discussed.

\end{abstract}
\maketitle


\section{Introduction  
\label{sec:Intro} 
}
\noindent
Electromagnetic (EM) interactions serve as valuable probes by which to shed light on the internal structure of strongly interacting single and multi-hadron systems. They provide insight into the charge and current distributions inside the hadrons. These are conventionally characterized by EM form factors, and are accessible through experimental measurements of electron-hadron scattering as well as EM transitions. The static and quasi-static limits of form factors, known as EM moments and charge radii, are independently accessible through high-precision low-energy experiments, such as in the spectroscopy of electronic and muonic atoms. These two different experimental approaches can serve to test the accuracy of the obtained quantities, and an apparent discrepancy, such as the one reported on the charge radius of the proton \cite{Pohl:2013, Carlson:2015jba}, promotes investigations that can deepen our understanding of the underlying dynamics. In bound systems of nucleons, EM probes further serve as a tool to constrain the form of hadronic forces. As a primary example, the measurement of a nonvanishing electric quadrupole moment for the lightest nucleus, the deuteron, led to the establishment of the existence of tensor components in the nuclear forces \cite{PhysRev.57.677}.

Since quantum chromodynamics (QCD) governs the interactions of quark and gluon constituents of hadrons, any theoretical determination of the EM properties of hadronic systems must tie to  a QCD description. The spread of theoretical predictions based on QCD-inspired models, such as those reported on the EM moments of vector mesons \cite{Aliev:2004uj, Braguta:2004kx, Choi:2004ww, Bhagwat:2006pu}, highlights the importance of performing first-principles calculations that only incorporate the parameters of quantum electrodynamics (QED) and QCD as input. The only such calculations are those based on the method of lattice QCD (LQCD), and involve a numerical evaluation of the QCD path integral on a finite, discrete spacetime. By controlling/quantifying the associated systematics of these calculations, the QCD values of hadronic quantities can be obtained with systematically improvable uncertainties.

QED can be introduced in LQCD calculations, alongside with QCD, in the generation of gauge-field configurations. This, however, leads to large finite-volume (FV) effects arising from the long range of QED interactions \cite{Hayakawa:2008an, Davoudi:2014qua, Borsanyi:2014jba, Endres:2015gda, Lucini:2015hfa}. The numerical cost of a lattice calculation which treats photons as dynamical degrees of freedom has forbidden comprehensive first-principles studies of EM properties of hadrons and nuclei through this avenue.\footnote{Significant progress has been made in recent years on this front, resulting in increasingly more precise determinations of QED corrections to mass splittings among hadronic multiplets \cite{Blum:2007cy,Basak:2008na,Blum:2010ym,Portelli:2010yn,Portelli:2012pn,Aoki:2012st,deDivitiis:2013xla,Borsanyi:2013lga,Drury:2013sfa, Borsanyi:2014jba, Endres:2015gda, Horsley:2015vla}, and recently more refined calculations of the hadronic light-by-light contribution to the muon anomalous magnetic moment, albeit at unphysical kinematics \cite{Blum:2014oka}.} Alternatively, as is done in most studies of hadron structure, the matrix elements of the EM currents can be accessed through the evaluation of three-point correlation functions in a background of pure QCD gauge fields, with insertions of quark-level current operators  between hadronic states.

An alternative method, that has advantages over the aforementioned methods with regard to its simplicity, and potentially its computational costs, is the background field method. In this approach, a background EM field can be introduced in a LQCD calculation by imposing the $U(1)$ gauge links onto the $SU(3)$ gauge links.\footnote{In order to reduce the computational cost, one may introduce the $U(1)$ gauge links solely in the valence-quark sector of QCD. With this approximation, one can only reliably study those EM properties of the state that do not receive contributions from the sea quarks (receive no \emph{disconnected} contributions).} This is motivated by original experimental determinations of the static EM properties of hadrons and nuclei in external EM fields. By measuring the difference in the energy of the system with and without the background fields, and by matching to the knowledge of the Hamiltonian of the system deduced from the appropriate effective hadronic theory \cite{Caswell:1985ui, Labelle:1992hd, Labelle:1997uw, Kaplan:1998tg, Kaplan:1998we, Chen:1999tn, Detmold:2004qn, Detmold:2006vu, Detmold:2009dx, Hill:2012rh, Lee:2013lxa, Lee:2014iha}, the parameters of the low-energy Hamiltonian, i.e., those characterizing the coupling of the composite hadron to external fields, can be systematically constrained. This procedure has been successfully implemented to determine the magnetic moments of single hadrons and their electric and magnetic polarizabilities \cite{Bernard:1982yu, Martinelli:1982cb, Fiebig:1988en, Christensen:2004ca, Lee:2005ds, Lee:2005dq, Detmold:2006vu, Aubin:2008qp, Detmold:2009dx, Detmold:2010ts, Primer:2013pva, Lujan:2014kia}. The utility of this method in accessing information about the structure of nuclei has been demonstrated recently through a determination of the magnetic moments and polarzabilities of nuclei with atomic number $A<5$ (at an unphysically heavy light-quark mass) \cite{Beane:2014ora, Chang:2015qxa}. It is desirable to gain further insight into the structure of these nuclei by studying their charge radii and quadrupole moments (for nuclei with spin $\geq 1$). These quantities require new developments that extend the implementation of uniform background fields to the case of \emph{nonuniform} fields. We have recently presented such developments in Ref. \cite{Davoudi:2015cba}, providing the recipe for implementing general nonuniform background fields that satisfy the periodicity of a FV calculation.\footnote{See  Refs. \cite{Lee:2011gz, Engelhardt:2011qq} for previous implementations of selected nonuniform, but nonperiodic, background EM fields in LQCD calculations of spin polarizabilities of the nucleon, and Ref. \cite{Bali:2015msa} for a periodic implementation of a plane-wave EM field in a LQCD calculation of the hadronic vacuum polarization function.} In the present paper, motivated by interest in extracting the quadrupole moment of the deuteron, we provide the theoretical framework for performing a systematic matching between a suitable hadronic theory for spin-1 fields and the corresponding LQCD calculations in nonuniform background fields. Although LQCD studies of partial-wave mixing in the ${^3}S_1-{^3}D_1$ two-nucleon coupled channel can also reveal the noncentral feature of nuclear forces as demonstrated in Refs. \cite{Briceno:2013lba, Briceno:2013bda, Orginos:2015aya}, only a direct evaluation can incorporate the short-distance contributions to the quadrupole moment \cite{Kaplan:1998sz, Chen:1999tn}.\footnote{Here we must distinguish the mass quadrupole moment of the deuteron from its electric quadrupole moment. It is the former that may be related to the S/D mixing in the deuteron channel. Although these two moments are comparable in the physical world, this might not be the case necessarily at unphysical values of quark masses.} Other phenomenologically interesting quantities such as the (electric and magnetic) charge radii, which have been calculated so far through studies of the momentum dependence of the form factors,\footnote{See Ref. \cite{deDivitiis:2012vs} for an alternative method to extract the form factors at zero momentum transfer by evaluating the derivatives of the correlation functions with respect to external momenta. This method circumvents the need for an extrapolation to zero momentum transfer, and has been extended in Ref. \cite{Tiburzi:2014yra} to access the charge radii.} can be also accessed via the nonuniform field technique. This formalism is equally applicable to the case of scalar and vector mesons so as long as they are nearly stable with regard to strong and EM interactions.\footnote{This assumption remains justified for several vector resonances such as the $\rho$ meson at heavy quark masses.}

In Sec. \ref{sec:Rel-EFT}, we present a general effective field theory (EFT) of composite vector particles coupled to perturbatively weak EM fields. Such effective theories have been worked out extensively in both classic and modern literature, with features and results that sometimes differ one another. Here we follow the most natural path, building up the Lagrangian of the theory from the most general set of nonminimal interactions (those arising from the composite nature of the fields) consistent with symmetries of the relativistically covariant theory, in an expansion in $1/M$. $M$ denotes a typical scale of the hadronic theory which we take to be the physical mass of the composite particle. Since the organization of nonminimal couplings is only possible in the low-energy limit, this approach, despite its relativistically covariant formulation, can only be considered to be \emph{semi-relativistic}. This means that the spin-1 field satisfies a relativistic dispersion relation in the absence of EM fields. However, once these external fields are introduced, one only accounts for those nonminimal interactions that will be relevant in the nonrelativistic (NR) Hamiltonian of the system at a given order in $\frac{1}{M}$ expansion (see Refs. \cite{Lee:2013lxa, Lee:2014iha} for a similar strategy in the case of spin-0 and spin-$\frac{1}{2}$ fields). We next match the low-energy parameters of the semi-relativistic Lagrangian to on-shell processes at low momentum transfers, and discuss subtleties when electromagnetism is only introduced through  classical fields. The effective theory developed here relies on a $10$-component representation of the vector fields which reveals a first order (with respect to time derivative) set of equations of motion (EOM). It resembles largely that presented in earlier literature by Sakata and Taketani \cite{Taketani:1940}, Young and Bludman \cite{Young:1963zza}, and Case \cite{PhysRev.95.1323}, but has also new features. In particular, it incorporates the most general nonminimal couplings at $\mathcal{O}\left(\frac{1}{M^2}\right)$ and therefore systematically includes operators that probe the electric and magnetic charge radii of the composite particle. The semi-relativistic Green's functions are then constructed in Sec. \ref{sec:Rel-Greens-functions} for the case an electric field varying linearly in a spatial direction. These Green's functions are related to the quantum-mechanical propagator of anharmonic oscillator and have no closed analytics forms, making it complicated to match them to LQCD calculations.

To match to lattice correlation functions, it is of practical convenience to first deduce an effective NR Hamiltonian via the standard procedure of Foldy, Wouthuysen and Case \cite{Foldy:1949wa, PhysRev.95.1323}, as presented in Sec. \ref{sec:EOM}. We derive the quantum-mechanical wavefunctions of spin-1 particles in a linearly varying electric field (in space) and their corresponding Green's functions in Sec. \ref{sec:NR-Greens-functions}, and show that, for a particular choice of the field, they are the Landau-level wavefunctions of a particle trapped in a harmonic potential. Despite their simple form, these NR Green's functions can not be directly matched to LQCD correlation functions, unless a NR transformation is performed on the correlation functions, or alternatively, an inverse transformation is applied to NR Green's functions, as discussed in Sec. \ref{subsec:NR-transformed-correlators}. This leads to at least two practical strategies to constrain the EM couplings of the low-energy theory, namely the quadrupole moment and the electric charge radius, as are presented in Secs. \ref{subsec:NR-transformed-correlators} and \ref{subsec:Landau-projection}: one may try to match the transformed correlation function to the NR Green's function directly, or alternatively, by projecting the NR Green's functions onto given Landau eigenstates to identify the NR energy eigenvalues, and match them to the NR limit of energies extracted from the long-(Euclidean) time behavior of (spatially projected) LQCD correlation functions. Finally, the extracted quadrupole moment and charge radius must be extrapolated to their infinite-volume values by performing calculations in multiple volumes, or by determining their volume dependencies through an effective theory that is sensitive to the substructure of the hadron or nuclei, (e.g., chiral perturbation theory in the former case and pionless EFT in the latter). By inputting the knowledge of the charge radius and the quadrupole moment of the deuteron, we have investigated the range of validity of the results obtained in this paper under a single-particle effective theory of the deuteron given various electric field choices. The viability of an extraction of the deuteron's quadruple moment and the charge radius within the framework of this paper from future LQCD calculations is then discussed, as presented in Sec. \ref{subsec:Deuteron}. We conclude in Sec. \ref{sec:Conclusion} by summarizing the results and commenting on future extensions. Additionally, the paper includes two appendices: appendix \ref{App:Gauge-dependency} is devoted to clarify the gauge dependency of the relativistic Green's functions of Sec. \ref{sec:Rel-Greens-functions}, and Appendix \ref{App:Rel-NR-relation} discusses the relation between the relativistic and NR Green's functions through an example.

\section{Composite Spin-1 Particles Coupled to External Electromagnetic Fields
\label{sec:Rel-EFT} 
}
\noindent
Any relativistic description of massive vector particles, due to the requirement of Lorentz invariance, must introduce fields that have redundant degrees of freedom. The most obvious choice is to represent the spin-1 field by a Lorentz four-vector, $V^{\mu}$, the so-called  Proca field \cite{Proca:1936}. The redundant degree of freedom of the Proca field, $V^0$, can  be eliminated using the EOM. These EOM are second order differential equations, and their reduced form, i.e., after the elimination of the redundant component, turns out to be non-Hermitian. Consequently, the solutions are in general nonorthogonal and difficult to construct in external EM fields \cite{Silenko:2004ww}. To avoid these difficulties, an equivalent formalism can be adopted by casting the Proca equation into coupled first-order differential equations, known as the Duffin-Kemmer equations \cite{Duffin:1938zz, Kemmer:1939zz}. This requires raising the number of degrees of freedom of the field and consequently introducing more redundancies. However, these redundant components can be eliminated in a straightforward manner, leading to EOM that can be readily solved (see the next section). There is a rich literature on relativistic spin-1 fields and their couplings to external EM fields via different first- and second-order formalisms, see for example Refs. \cite{Corben:1940zz, Vijayalakshmi:1978ju, Santos:1986yw, Daicic:1993, Khriplovich:1997ni, Pomeransky:1999ej, Silenko:2004ww, Silenko:2006pj, Silenko:2013oya}. Here we follow closely the work of Young and Bludman \cite{Young:1963zza} which is  a generalization of first-order Sakata-Taketani equations for spin-1 fields \cite{Taketani:1940}. However, due to the spread of existing results, and occasionally inconsistencies among them, we independently work out the construction of an EFT for massive spin-$1$ fields towards our goal of deducing Green's functions of spin-1 fields in a selected external field. In particular, the nonminimal couplings in our Lagrangian, as will be discussed shortly, are more general than those presented in all previous studies, and include all the possible terms needed to consistently match to not only the particle's electric quadrupole moment but also its electric and magnetic charge radii at $\mathcal{O}\left(\frac{1}{M^2}\right)$ (we neglect terms that are proportional to the field-strength squared with coefficients that are matched to polarizabilities). Although fields and interactions have been described in a Lorentz-covariant relativistic framework, the nonminimal couplings to external fields can only be organized in an expansion in the mass of the particle, or in turn a generic hadronic scale above which the single-particle description breaks down.\footnote{Although the expansion parameter is taken to be the mass, the size of nonminimal interactions is indeed governed by the compositeness scale of the particle. In fact, as we will see shortly, when these compositeness scales, such as radii and moments, arise in matching the coefficients to on-shell processes, the factors of mass cancel.} At low energies, one can truncate these nonminimal interactions at an order such that, after a full NR reduction, the effective theory incorporates information about as many low-energy parameters as one is interested in.

\subsection{A semi-relativistic effective field theory
\label{subsec:semi-rel-EFT}}
We start by writing down the most general Lorentz-invariant Lagrangian for a single massive spin-1 field, coupled to electromagnetism, that is invariant under charge conjugation, time reversal and parity. We choose to construct the Lagrangian out of a four-component field $V^{\alpha}~(\alpha=0,1,2,3)$ and a rank-two tensor $W^{\mu \nu}$ ($\mu,\nu=0,1,2,3$). However, as we shall see below, the EOM of the resulting theory constrain the number of independent degrees of freedom to those needed to describe the physical modes of a spin-1 field. The Lagrangian, in terms of $V^{\alpha}$ and $W^{\mu \nu}$ degrees of freedom, can be written as
\begin{eqnarray}
\mathcal{L}=\frac{1}{2}W^{\dagger \mu \nu}W_{\mu \nu}+
M^2V^{\dagger \alpha}V_{\alpha}
-\frac{1}{2}W^{\dagger \mu \nu}(D_{\mu}V_{\nu}-D_{\nu}V_{\mu})
-\frac{1}{2}((D_{\mu}V_{\nu})^{\dagger}-D_{\nu}V^{\dagger}_{\mu})W^{\mu \nu}+ \qquad \qquad \qquad
\nonumber\\
ieC^{(0)}~F_{\mu \nu}~V^{\dagger \mu}V^{\nu}+
\frac{ieC^{(2)}_1}{M^2}~\partial_{\mu}F^{\mu \nu} ((D_{\nu}V^{\alpha})^{\dagger} V_{\alpha}-V^{\dagger \alpha} D_{\nu}V_{\alpha})+ \qquad \qquad \qquad \qquad \qquad \qquad ~~~~~~~~
\nonumber\\
\frac{ieC^{(2)}_2}{M^2} ~\partial^{\alpha}F^{\mu \nu} ((D_{\alpha}V_{\mu})^{\dagger} V_{\nu}-V^{\dagger}_{\nu} D_{\alpha}V_{\mu})+
\frac{ieC^{(2)}_3}{M^2} ~\partial^{\nu}F^{\mu \alpha} ((D_{\mu}V_{\alpha})^{\dagger} V_{\nu}-V^{\dagger}_{\nu} D_{\mu}V_{\alpha})+
\mathcal{O}\left(\frac{1}{M^4},F^2\right), ~
\label{eq:Lagrangian-Rel-6}
\end{eqnarray}
where $D_{\mu}=\partial_{\mu}+ieQ_0A_{\mu}$ denotes the covariant derivate, $F_{\mu \nu}=\partial_{\mu}A_{\nu}-\partial_{\nu}A_{\mu}$ is the EM field strength tensor, $A^{\mu}$ denotes the photon gauge field, and $Q_0$ refers to the electric charge of the particle. The superscripts on the coefficients denote the order of the corresponding terms in an expansion in $\frac{1}{M}$. By $\mathcal{O}\left(\frac{1}{M^4}\right)$ we indicate any Lorentz-invariant term bilinear in $V/W$ and $V^{\dagger}/W^{\dagger}$ with appropriate numbers of covariant derivates and $F^{\mu\nu}$s such that the overall mass dimension is four when accompanied by $\frac{1}{M^4}$. Similarly, $\mathcal{O}\left(F^2\right)$ corresponds to any Lorentz-invariant term with mass dimension four that contains two $F^{\mu \nu}$s. In particular, this latter include $\frac{1}{M^2}F^2V^{\dagger}V$-type interactions that are of the same order in the inverse mass expansion as are the nonminimal terms we have considered, and whose coefficients are matched to electric and magnetic polarizabilities of the particle. By assuming a small external field strength, we can neglect these contributions. In order to access polarizabilities, Eq. (\ref{eq:Lagrangian-Rel-6}) must be revisited to include such terms.

The coefficients of the leading contributions are fixed to reproduce the canonical normalization of the resulting kinetic term for massive spin-1 particles \cite{Proca:1936}. We have taken advantage of the following property of the EM field strength tensor $\partial_{\nu}F_{\mu \alpha}=\partial_{\mu}F_{\nu \alpha}+\partial_{\alpha}F_{\mu \nu}$ to eliminate redundant terms at $\mathcal{O}\left(\frac{1}{M^2}\right)$. Additionally, the number of terms with a given Lorentz structure at each order can be considerably reduced by using the constraint of vanishing surface terms in the action. This constraint is not trivial in the presence of EM background fields which extend to infinite boundaries of spacetime (which is an unphysical but technically convenient situation). To rigorously define a field theory in the background of classical fields, one shall assume background fields are finite range, are adiabatically turned on in distant past and will be adiabatically turned off in far future. Mathematically, this means that one must accompany external fields by a factor of $e^{-\eta^{\mu}|x_{\mu}|}$, where $\eta^{\mu}$ is positive and $\eta^{\mu} \rightarrow 0$. This ensures that for any finite value of $x_{\mu}$, the background field is $\eta$ independent and nonzero, while as $x_{\mu} \rightarrow \pm \infty$, the field gradually vanishes. This procedure is particularly important when space-time dependent background fields are considered. This is because the sensibility of the expansion of nonminimal couplings in Eq. (\ref{eq:Lagrangian-Rel-6}) when $x_{\mu} \rightarrow \pm \infty$ is guaranteed only if a mechanism similar to what described above is in place. In a calculation performed in a finite volume, such a procedure does not eliminate the contributions at the boundary.  However, in this case one is free to choose the boundary conditions. For example, if periodic boundary conditions (PBCs) are imposed on the fields, the contributions of the surface terms to the action will in fact vanish just as in the infinite volume. As a result, the only relevant interactions in both scenarios have been already included in the Lagrangian in Eq. (\ref{eq:Lagrangian-Rel-6}), with coefficients that could be meaningfully constrained by matching to on-shell processes in the infinite spacetime volume. To satisfy PBCs in a finite volume, certain quantization conditions must be imposed on the parameters of the background fields, which can be seen to also prevent potential large background field strengths at the boundaries of the volume, see Ref. \cite{Davoudi:2015cba}.

The Euler-Lagrange EOM arising from the Lagrangian in Eq. (\ref{eq:Lagrangian-Rel-6}) are
\begin{eqnarray}
&&\text{(I)}~~~W_{\mu \nu}=D_{\mu}V_{\nu}-D_{\nu}V_{\mu}+\mathcal{O}\left(\frac{1}{M^4},F^2\right),
\label{eq:EOM-Rel-6-comp-I}
\end{eqnarray}
\begin{eqnarray}
&&\text{(II)}~~~D^{\mu}W_{\mu \alpha}+
M^2V_{\alpha}+ieQ_0C^{(0)}F_{\alpha \mu}V^{\mu}=
\frac{ie}{M^2} \left[ 2C^{(2)}_1\partial_{\mu}F^{\mu \nu}D_{\nu}V_{\alpha}+ C^{(2)}_2 \partial^2F_{\alpha \nu}V^{\nu}+ \right .
\nonumber\\
&& \qquad \qquad \qquad \qquad \left . C^{(2)}_3\left(\partial_{\nu}F_{\mu \alpha}D^{\mu}V^{\nu}+\partial_{\alpha}F_{\mu \nu}D^{\mu}V^{\nu}+\partial^{\mu}\partial^{\nu}F_{\mu \alpha}V_{\nu}\right) \right] +\mathcal{O}\left(\frac{1}{M^4},F^2\right),
\label{eq:EOM-Rel-6-comp-II}
\end{eqnarray}
where $\mathcal{O}\left(\frac{1}{M^4}\right)$ in Eq. (\ref{eq:EOM-Rel-6-comp-I}) (Eq. (\ref{eq:EOM-Rel-6-comp-II})) denotes any Lorentz-invariant term with mass dimension two (three) with at most one $V$ or $W$ field. Similarly, $\mathcal{O}\left(F^2\right)$ in Eq. (\ref{eq:EOM-Rel-6-comp-I}) (Eq. (\ref{eq:EOM-Rel-6-comp-II})) denotes any Lorentz-invariant terms with mass dimension two (three) with at least two powers of the field strength tensor and at most one $V$ or $W$ field. Note that from the first equation, it is established that $W^{\mu \nu}$ is an antisymmetric tensor up to $\mathcal{O}\left(\frac{1}{M^4},F^2\right)$ corrections. We have anticipated this feature in writing down all possible terms at $\mathcal{O}\left(\frac{1}{M^2}\right)$ in the Lagrangian Eq. (\ref{eq:Lagrangian-Rel-6}), as the nonantisymmetric piece of $W^{\mu \nu}$ gives rise to contributions that are of higher orders. This also makes any term containing one $W^{\mu \nu}$ and one $V^{\mu}$ field at $\mathcal{O}\left(\frac{1}{M^2}\right)$ redundant.

In writing the Lagrangian in Eq. (\ref{eq:Lagrangian-Rel-6}), we have neglected terms of the type $\frac{1}{M^2}V^{\dagger}D^4V$. These can be reduced to terms that have been already included in the Lagrangian at this order using the EOM. A number of inconsistencies might occur when the EOM operators are naively discarded in the presence of background fields. However, as is discussed in Refs. \cite{Lee:2013lxa, Lee:2014iha}, the neglected terms in the Lagrangian only modify Green's functions by overall spacetime-independent factors that can be safely neglected. The other sets of operators at $\mathcal{O}\left(\frac{1}{M^2}\right)$ that we have taken the liberty to exclude due to the constraint from the EOM are those containing at least one $D_{\mu}V^{(\dagger)\mu}$. These vanish up to corrections that scale as $\mathcal{O}\left(\frac{F}{M^2}\right)$ (see Eqs. (\ref{eq:EOM-Rel-6-comp-I}) and (\ref{eq:EOM-Rel-6-comp-II}) above), and therefore give rise to higher order terms, i.e., $\mathcal{O}\left(\frac{1}{M^4},F^2\right)$, in the Lagrangian.\footnote{According to Refs. \cite{Lee:2013lxa, Lee:2014iha}, the EOM operators in fact must be given special care only in the \emph{NR} theory. The contribution from these operators to on-shell processes could be nontrivial in situations where QED is introduced through a background EM field. Given that we follow a direct NR reduction of the relativistic theory, all such subtleties will be automatically taken care of. In particular, it is notable that the semi-relativistic Lagrangian with a background electric field up to $\mathcal{O}\left(\frac{1}{M^2}\right)$ generates terms of the type $\frac{F^2}{M^3}$ in the NR Hamiltonian, see Sec. \ref{sec:EOM}. This is despite the fact that we have already neglected terms of $\mathcal{O}\left(\frac{F^2}{M^2}\right)$ in the semi-relativistic Lagrangian. These are the type of contributions that are shown to correspond to an EOM operator in the scalar NR Lagrangian, and will add to contributions that correspond to a polarizability shift in the energy of the NR particle. It is shown in Refs. \cite{Lee:2013lxa, Lee:2014iha} that by keeping track of these terms, inconsistencies that are observed in the second-order energy shifts of spin-0 and spin-$\frac{1}{2}$ particles in uniform external electric fields can be resolved. Although we do not explicitly work out the polarizability contributions in this paper, we expect the same mechanism to be in place with our framework for the case of spin-1 fields.}

Before concluding the discussion of the semi-relativistic Lagrangian, it is worth pointing out that a number of pathologies have been noted in literature for relativistic theories of massive spin-1 (and higher) particles in background (EM or gravitational) fields. One issue that is most relevant to our discussion here is the emergence of superluminal modes from nonminimal couplings (such as quadrupole coupling) to EM fields, as noted by Velo and Zwanziger \cite{Velo:1970ur}. However, as is discussed in Ref. \cite{Porrati:2008gv}, the acasuality arising from nonminimal interactions are manifest as singularities (that can not be removed by any field redefinition) when one takes the $M \rightarrow 0$ limit. Therefore, the pathologies associated with these modes arise at a scale which is comparable or higher than the mass of the vector particle. Since the effective theory for nonminimal couplings already assumes a cutoff scale of $\sim M$, these pathologies are not relevant in our discussions. Thus, there in no contradiction to the existence of a well-defined low-energy effective theory that describes interactions of particles with any spin in external fields, as characterized by their EM moments, polarizabilities, and their higher static and quasi-static properties. With the assumption of weak external EM fields, other possibilities discussed in literature, such as the spontaneous EM superconductivity of vacuum due to the charged vector-particle condensation \cite{Ambjorn:1988tm, Ambjorn:1988gb, Chernodub:2011mc}, will not be relevant in the framework of this paper.

\
\

In what follows, we carry out the matching to on-shell amplitudes at low-momentum transfer to constrain the values of the coefficients in the effective Lagrangian.

\subsection{Matching the effective theory to on-shell amplitudes
}
\emph{Electromagnetic current and form-factor decomposition}: The form-factor decomposition of the matrix elements of the EM current for spin-1 particles is well known, as is its connection to the EM multipole decomposition of NR charge and current densities, see for example Refs. \cite{Arnold:1979cg, Lorce:2009br}.  We briefly review the relevant discussions; this also serves as an introduction to our conventions.

Considering Lorentz invariance, vector-current conservation and charge-conjugation invariance, the most general form of the matrix element of an EM current, $J^{\mu}$, between on-shell vector particles can be written as 
\begin{eqnarray}
\bra{p',\lambda'}J^{\mu}(q)\ket{p,\lambda}= -e~{\epsilon^{(\lambda')}_{\alpha}(p')}^{\dagger} \left [ F_1(Q^2) P^{\mu}g^{\alpha \beta}+
F_2(Q^2)(g^{\mu \beta}q^{\alpha}-g^{\mu \alpha}q^{\beta}) -\frac{F_3(Q^2)}{2M^2} q^{\alpha}q^{\beta}P^{\mu} \right] \epsilon^{(\lambda)}_{\beta}(p),
\nonumber\\
\label{eq:current-decomp}
\end{eqnarray}
where $\ket{p,\lambda}$ denotes the initial state of a vector particle with momentum $p$ and polarization $\lambda$, and $\bra{p',\lambda'}$ denotes its final state with momentum $p'$ and polarization $\lambda'$, and where the momentum transferred to the final state due to interaction with the EM current is $q=p'-p$. $\epsilon^{(\lambda)}(p)$ denotes the $\lambda^{\text{th}}$ polarization vector of the particle with momentum $p$. For massive on-shell particles $\lambda$ runs from $1$ to $3$. Additionally, $P=p+p'$ and we have defined $Q^2=-q^2$. Lorentz structures proportional to $P^{\mu}P^{\alpha}P^{\beta}$, $P^{\mu}(P^{\alpha}q^{\beta}-q^{\alpha}P^{\beta})$ and $q^{\mu}(P^{\alpha}q^{\beta}+q^{\alpha}P^{\beta})$ have been discarded by utilizing the following conditions on the polarization vectors: ${p'}^{\alpha}\epsilon^{(\lambda')}_{\alpha}(p')=0$ and $p^{\beta}\epsilon^{(\lambda)}_{\beta}(p)=0$. Although the right-hand sides of these conditions are modified in external electric, $\bm{E}$, and magnetic, $\bm{B}$, fields by terms of $\mathcal{O}\left( \frac{\bm{E}}{M^2},\frac{\bm{B}}{M^2} \right)$, this will not matter for calculating on-shell matrix element as long as the adiabatic procedure described above Eq. (\ref{eq:EOM-Rel-6-comp-I}) is in place to eliminate surface terms in the Lagrangian. By introducing the external fields adiabatically, the asymptotic ``in'' and ``out'' states of the theory are free and the corresponding polarization vectors satisfy the noninteracting relations.

To relate the form factors in Eq. (\ref{eq:current-decomp}) at low $Q^2$ to the low-energy EM properties of the spin-1 particle, one may interpret this current matrix element, when expressed in the Breit frame, as multipole decomposition of the classical electric and magnetic charge densities. These decompositions are defined through Sachs form factors,
\begin{eqnarray}
\rho_E(\mathbf{q})   \equiv   \int d^3x e^{i\mathbf{q}\cdot\bm{x}} J_{cl}^{0}(\bm{x})
= e \sum_{l=0,~l~\text{even}}^{2S}\left(-\frac{Q^2}{4M^2}\right)^{\frac{l}{2}}\sqrt{\frac{4\pi}{2l+1}}\frac{l!}{(2l-1)!!} G_{El}(Q^2)Y_{l0}(\hat{0}),
\qquad \qquad 
\label{eq:rho}
\end{eqnarray}
\begin{eqnarray}
\rho_M(\mathbf{q})  \equiv  \int d^3x e^{i\mathbf{q}\cdot\bm{x}}  \bm{\nabla} \cdot \left(\bm{x} \times\mathbf{J}_{cl}(\bm{x})\right)
= e \sum_{l=0,~l~\text{odd}}^{2S}  \left(-\frac{Q^2}{4M^2}\right)^{\frac{l}{2}}\sqrt{\frac{4\pi}{2l+1}}\frac{(l+1)l!}{(2l-1)!!} G_{Ml}(Q^2)Y_{l0}(\hat{0}),
\label{eq:rho-M}
\end{eqnarray}
where $G_{El}$ and $G_{Ml}$ are the $l^{\text{th}}$ Sachs electric and magnetic form factors, respectively, and $S$ denotes the value of spin. If the particle was infinitely massive, such interpretation of the relativistic relation (\ref{eq:current-decomp}) would have been exact, and the current matrix element would be precisely the Fourier transform of some classical charge or current density distributed inside the hadron. However, away from this limit, there are small recoil effects at low energies that are hard to characterize in the hadronic theory. In the Breit frame, in which the energy of the transferred photon, $q^0$, is zero, such effects are minimal as the initial and final states have the same energy. In fact, as is well known, by expressing Eq. (\ref{eq:current-decomp}) in this frame, and by taking the moving-frame polarizations vectors satisfying  ${p'}^{\alpha}\epsilon^{(\lambda')}_{\alpha}(p')=0$ and $p^{\beta}\epsilon^{(\lambda)}_{\beta}(p)=0$,
this matrix element resembles the classical forms in Eqs. (\ref{eq:rho}) and (\ref{eq:rho-M}). This enables one to directly relate the form factors $F_1(Q^2),~F_2(Q^2)$
 and $F_3(Q^2)$, to Sachs form factors $G_{E_0}(Q^2),~G_{E_2}(Q^2)$ and $G_{M_1}(Q^2)$. For spin-$1$ particles this results in the relations
\begin{eqnarray}
G_{E0}(Q^2) & \equiv & G_{C}(Q^2) = F_1(Q^2)+\frac{2}{3}\frac{Q^2}{4M^2}G_{E2}(Q^2),
\\
G_{E2}(Q^2) & \equiv & G_{Q}(Q^2) = F_1(Q^2)-F_2(Q^2)+(1+\frac{Q^2}{4M^2})F_3(Q^2),
\\
G_{M1}(Q^2) & \equiv & G_{M}(Q^2) = F_2(Q^2).
\label{eq:form-factors}
\end{eqnarray}

The electric charge, electric quadrupole moment and magnetic dipole moment are defined as the zero momentum transfer limit of the Coulomb, $G_{C}(Q^2)$, quadrupole, $G_{Q}(Q^2)$, and magnetic, $G_{M}(Q^2)$, Sachs form factors, respectively,
\begin{eqnarray}
\label{eq:moments-I}
Q_0 & \equiv & G_{C}(0) = F_1(0),
\\
\label{eq:moments-II}
\overline{Q}_2 & \equiv & G_{Q}(0) = F_1(0)-F_2(0)+F_3(0),
\\
\overline{\mu}_1 & \equiv & G_{M}(0) = F_2(0).
\label{eq:moments-III}
\end{eqnarray}
$\overline{Q}_2$ is the particle's quadrupole moment in units of $\frac{e}{M^2}$, and $\overline{\mu}_1$ denotes its magnetic moment in units of $\frac{e}{2M}$.
Additionally, the mean-squared electric and magnetic charge radii can be expressed, respectively, as the derivatives of the Coulomb and magnetic form factors with respect to $Q^2$ at zero momentum transfer,
\begin{eqnarray}
\label{eq:radius-E}
\braket{r^2}_E & \equiv & -6e \left . \frac{dG_C(Q^2)}{dQ^2} \right |_{Q^2=0} 
=  -6e\left . \frac{dF_1(Q^2)}{dQ^2} \right |_{Q^2=0}-\frac{e\overline{Q}_2}{M^2},
\\
\braket{r^2}_M & \equiv & -6e \left . \frac{dG_M(Q^2)}{dQ^2} \right |_{Q^2=0} = -6e \left . \frac{dF_2(Q^2)}{dQ^2} \right |_{Q^2=0}.
\label{eq:radius-M}
\end{eqnarray}
The quadrupole charge radius can be defined similarly from the derivative of the quadrupole Sachs form factor, however the dependence on this radius only occurs at higher orders in $\frac{1}{M}$ than is considered below.

\
\

\emph{One-photon amplitude from the effective theory}: The next step is to evaluate the one-photon amplitude from the effective Lagrangian in Eq. (\ref{eq:Lagrangian-Rel-6}). Explicitly, the following quantity
\begin{eqnarray}
\Gamma^{\alpha \beta \mu} \equiv -\bra{V^{\alpha}(p')}\mathcal{L}[V^{\dagger},V,W^{\dagger},W,A]\ket{V^{\beta}(p)A^{\mu}(q)},
\label{eq:EFT-ME-I}
\end{eqnarray}
must be evaluated from the Lagrangian in Eq. (\ref{eq:Lagrangian-Rel-6}) to match to Eq. (\ref{eq:current-decomp}). In obtaining this on-shell amplitude, the condition of the orthogonality of the momentum vectors to their corresponding polarization vectors can be used once again. Moreover, we use the EOM (see Eq. (\ref{eq:EOM-Rel-6-comp-I})) to convert $W^{\mu \nu}$ fields to $V^{\mu}$ fields. A straightforward but slightly lengthy calculation gives
\begin{eqnarray}
\Gamma^{\alpha \beta \mu}&=&-e \left \{  \left[Q_0+C^{(2)}_1\frac{q^2}{M^2}\right]g^{\alpha \beta} P^{\mu} -C_3^{(2)} \frac{ q^{\alpha}q^{\beta}}{M^2}P^{\mu}+ \right .
\nonumber\\
&& \qquad \qquad \qquad \qquad \left . \left[ Q_0+C^{(0)}+ (C_2^{(2)}-\frac{1}{2}C_3^{(2)})\frac{q^2}{M^2}\right] (g^{\mu \beta}q^{\alpha}-g^{\mu \alpha}q^{\beta}) \right \}.
\label{eq:EFT-ME-II}
\end{eqnarray}
By comparing Eqs. (\ref{eq:EFT-ME-II}) and (\ref{eq:current-decomp}), and with the aid of Eqs. (\ref{eq:moments-I})-(\ref{eq:radius-M}), the following relations can be deduced,
\begin{align}
& F_{1}(Q^2)= Q_0-C^{(2)}_1\frac{Q^2}{M^2}+\mathcal{O}\left(\frac{Q^4}{M^4}\right)=Q_0-\frac{1}{6e}\left(M^2 \braket{r^2}_E + e\overline{Q}_2 \right) \frac{Q^2}{M^2}+\mathcal{O}\left(\frac{Q^4}{M^4}\right),
\\
& F_{2}(Q^2)= Q_0+C^{(0)}-(C_2^{(2)}-\frac{1}{2}C_3^{(2)}) 
\frac{Q^2}{M^2}+\mathcal{O}\left(\frac{Q^4}{M^4}\right) 
= \overline{\mu}_1-\frac{M^2}{6e}\braket{r^2}_M \frac{Q^2}{M^2}+\mathcal{O}\left(\frac{Q^4}{M^4}\right),
\\
& F_{3}(Q^2)=2C_3^{(2)} +\mathcal{O}\left(\frac{Q^2}{M^2}\right)
= (-Q_0+\overline{Q}_2+\overline{\mu}_1)+\mathcal{O}\left(\frac{Q^2}{M^2}\right).
& \label{eq:matching-Rel}
\end{align}
These fully constrain the values of the four coefficients in the effective Lagrangian as following
\begin{eqnarray}
\label{eq:LECs-Rel-I}
&& C^{(0)}= \overline{\mu}_1-Q_0,
\\
\label{eq:LECs-Rel-II}
&& C^{(2)}_1= \frac{1}{6e}\left(M^2 \braket{r^2}_E + e\overline{Q}_2 \right),
\\
\label{eq:LECs-Rel-III}
&& C_{2}^{(2)} = \frac{1}{4}(-Q_0+\overline{Q}_2+\overline{\mu}_1)+\frac{1}{6e}M^2 \braket{r^2}_M,
\\
\label{eq:LECs-Rel-IV}
&& C_3^{(2)} = \frac{1}{2}(-Q_0+\overline{Q}_2+\overline{\mu}_1).
\end{eqnarray}

With nonminimal interactions being constrained by the on-shell amplitudes, Eq. (\ref{eq:Lagrangian-Rel-6}) can now be utilized to study properties of spin-1 particles in external fields. This is pursued in the next section through analyzing the EOM of the vector particle in time-independent but otherwise general $\bm{E}$ and $\bm{B}$ fields and their reduced forms in the NR limit.

\section{Equations of Motion in External Fields and their Nonrelativistic Reductions
\label{sec:EOM} 
}
\noindent
To be able to find the physical solutions of the EOM, one must first eliminate the redundant degrees of freedom of the spin-1 field in Eqs. (\ref{eq:EOM-Rel-6-comp-I}) and (\ref{eq:EOM-Rel-6-comp-II}). This can be established by eliminating $V^0$ and $W^{ij}$, with $i,j=1,2,3$, in favor of the remaining 6 components of the fields, namely
\begin{eqnarray}
V^i ~~ \text{and} ~~ \phi^i \equiv \frac{1}{M}W^{i0}.
\end{eqnarray}
Our choice here is justified by noting that these latter are the only dynamical components of the fields (according to Eqs. (\ref{eq:EOM-Rel-6-comp-I}) and (\ref{eq:EOM-Rel-6-comp-II}), the time derivatives of  $V^0$ and $W^{ij}$ are absent from the EOM).  From Eq. (\ref{eq:EOM-Rel-6-comp-I}) it is manifest that the $W^{ij}$ fields are related to the derivative of the $V^i$ fields
\begin{eqnarray}
W^{ij}=D^iV^j-D^jV^i.
\label{eq:Wij-replacement}
\end{eqnarray}
It is also deduced from Eq. (\ref{eq:EOM-Rel-6-comp-II}) that the $V^0$ field can be written in terms of the $\bm{V}$ and $\bm{\phi}$ fields,
\begin{eqnarray}
V^{0}=-\frac{\bm{D}\cdot \bm{\phi}}{M}-\frac{ieC^{(0)}}{M^2}\bm{E}\cdot \bm{V}+\mathcal{O}\left(\frac{1}{M^3}\right),
\label{eq:V0-replacement}
\end{eqnarray}
where $D^0=\frac{d}{dt}+ieQ_0\varphi$ and $\bm{D}=\bm{\nabla}-ieQ_0\bm{A}$. $\varphi$ and $\mathbf{A}$ refer to the scalar and vector EM potentials, respectively. The bold-faced quantities now represent ordinary three-vectors; as a result from here on we do not distinguish the upper and lower indices and let them all represent cartesian spatial indices. The terms that originate from the LHS of Eq. (\ref{eq:EOM-Rel-6-comp-II}) contribute to $V^0$ at $\mathcal{O}\left(\frac{1}{M^3}\right)$ or higher. As can be seen from the EOM for the dynamical fields (see below), such terms give rise to contributions that are of  $\mathcal{O}\left(\frac{1}{M^4}\right)$ or higher and will be neglected in our analysis. 
By taking into account these relations, and further by assuming time-independent external fields, the coupled EOM for the $\bm{V}$ and $\bm{\phi}$ fields can be written as
\begin{eqnarray}
&&i\frac{d\bm{\phi}}{dt}=M\bm{V}+eQ_0\varphi \bm{\phi}+\frac{1}{M}\bm{D} \times \bm{D} \times \bm{V}
-\frac{ieC^{(0)}}{M}\bm{B}\times \bm{V}+
\frac{eC^{(0)}}{M^2}\bm{E} \left(\bm{D} \cdot \bm{\phi}\right)
\nonumber\\
&& \hspace{0.25 cm}
-\frac{2eC^{(2)}_1}{M^2}\left[ (\overline{\bm{\nabla}} \cdot \bm{E}) \bm{\phi}+\frac{i}{M}\left(\overline{\bm{\nabla}}\times \bm{B} \cdot \bm{D}\right) \bm{V} \right]-\frac{ieC^{(2)}_2}{M^3}(\overline{\bm{\nabla}}^2\bm{B})\times \bm{V}+\frac{eC^{(2)}_3}{M^2}\left[ (\bm{\phi} \cdot \overline{\bm{\nabla}}) \bm{E}+ \right .
\nonumber\\
&& \hspace{0.25 cm}
\left . \overline{\bm{\nabla}} (\bm{E} \cdot \bm{\phi})+\frac{i}{M}\overline{\bm{\nabla}}(\bm{B} \times \bm{D}) \cdot \bm{V}
+\frac{i}{M}\overline{\bm{\nabla}}_k \left(\bm{B} \times \bm{D}\right) \bm{V}_k
-\frac{i}{M}(\bm{V} \cdot \overline{\bm{\nabla}})(\overline{\bm{\nabla}} \times \bm{B})  \right]+\mathcal{O}\left(\frac{1}{M^4},F^2\right),
\label{eq:phi-EOM-E-B}
\end{eqnarray}
\begin{eqnarray}
i\frac{d\bm{V}}{dt}=M\bm{\phi}+eQ_0\varphi \bm{V}-\frac{1}{M}\bm{D} (\bm{D} \cdot \bm{\phi})
-\frac{eC^{(0)}}{M^2}\bm{D} \left(\bm{E} \cdot \bm{V}\right)+\mathcal{O}\left(\frac{1}{M^4},F^2\right), 
\label{eq:V-EOM-E-B}
\end{eqnarray}
where we have transformed the $\bm{V}$ field to $-i\bm{V}$. The line over the derivatives indicates that the operator acts solely on the electric or magnetic field and not on the spin-1 fields following them.

These equations can be cast into an elegant matrix form. This can be achieved by introducing the following matrices
\begin{eqnarray}
S_1=\begin{pmatrix}
  0 & 0 & 0 \\
  0 & 0 & -i \\
  0 & i & 0
 \end{pmatrix},
 ~~~S_2=\begin{pmatrix}
  0 & 0 & i \\
  0 & 0 & 0 \\
  -i & 0 & 0
 \end{pmatrix},
  ~~~S_3=\begin{pmatrix}
  0 & -i & 0 \\
  i & 0 & 0 \\
  0 & 0 & 0
 \end{pmatrix},
\label{eq:spin-matrices}
\end{eqnarray}
with the properties: $S^2=S_1^2+S_2^2+S_3^2=2~ \mathbb{I}_{3 \times 3}$ and $[S_i,S_j]=i\epsilon_{ijk}S_k$, where $\epsilon_{ijk}$ is the three-dimensional Levi-Civita tensor. These matrices are closely related to the notion of spin in a NR theory as will become clear shortly.\footnote{These are the analogues of Pauli matrices for spin-$\frac{1}{2}$ particles.} In the following, the EOM are further analyzed by separating the case of electric and magnetic fields. This is solely to keep the presentation  tractable, and the results for the case of nonvanishing electric and magnetic fields can be straightforwardly obtained following the same procedure.

\subsection{An external electric field}
For the case of an electric field with no time variation, the EOM for the $\bm{V}$ and $\bm{\phi}$ fields can be rewritten as
\begin{eqnarray}
i\frac{d\bm{\phi}}{dt}=M\bm{V}+eQ_0\varphi \bm{\phi}-\frac{1}{M} (\bm{S} \cdot \bm{D})^2 \bm{V}+
\frac{eC^{(0)}}{M^2}\left[ \bm{E} \cdot \bm{D}-S_i S_j \bm{E}_j \bm{D}_i \right] \bm{\phi}
-\frac{2eC_1^{(2)}}{M^2}(\overline{\bm{\nabla}} \cdot \bm{E})\bm{\phi}+ \hspace{0.5 cm}
\nonumber\\
 \frac{2eC_3^{(2)}}{M^2} \left[ \overline{\bm{\nabla}} \cdot \bm{E} - \frac{1}{2}(S_i S_j+S_j S_i) \overline{\bm{\nabla}}_i \bm{E}_j \right] \bm{\phi}+
\mathcal{O}\left(\frac{1}{M^4},F^2\right),
\label{eq:EOM-phi-V-nonzero-E-I}
\end{eqnarray}
\begin{align}
i\frac{d\bm{V}}{dt}=M\bm{\phi}+eQ_0\varphi \bm{V}-\frac{1}{M} \left[ \bm{D}^2-(\bm{S} \cdot \bm{D})^2 \right] \bm{\phi}-\frac{eC^{(0)}}{M^2}\left[ \bm{D} \cdot \bm{E}-S_j S_i \bm{D}_i \bm{E}_j \right] \bm{V}+\mathcal{O}\left(\frac{1}{M^4},F^2\right),
\label{eq:EOM-phi-V-nonzero-E-II}
\end{align}
with the aid of spin matrices in Eq. (\ref{eq:spin-matrices}).
These two equations can be represented by a single EOM for a 6-component vector, conveniently defined as 
\begin{eqnarray}
\psi \equiv
\frac{1}{\sqrt{2}}
 \begin{pmatrix}
  \bm{\phi}+\bm{V} \\
  \bm{\phi}-\bm{V}
 \end{pmatrix}.
 \label{eq:psi-def}
\end{eqnarray} 
This equation resembles a Schr\"odinger equation for the field $\psi$,
\footnote{For this wavefunction, the expectation values of operators are defined by
\begin{eqnarray}
\overline{O}=\int d^3x \psi^{\dagger} \sigma_3 O \psi.
\end{eqnarray}
This imposes the condition of pseudo-Hermiticity on the Hamiltonian, $\widehat{\mathcal{H}}=\sigma_3\widehat{H}^{\dagger}\sigma_3$, which is clearly the case for the Hamiltonians in  Eqs. (\ref{eq:rel-Hamiltonian-E}) and (\ref{eq:rel-Hamiltonian-B}). See Ref. \cite{PhysRev.95.1323} for more details.}
\begin{eqnarray}
i\frac{d}{dt}\psi= \widehat{\mathcal{H}}^{(\bm{E})}_{\text{SR}} \psi,
\label{eq:EOM-psi-nonzero-E}
\end{eqnarray}
where the semi-relativistic Hamiltonian is
\begin{eqnarray}
\widehat{\mathcal{H}}^{(\bm{E})}_{\text{SR}}= M \sigma_3+eQ_0\widehat{\varphi}+(\sigma_3+i\sigma_2)\frac{\widehat{\bm{\pi}}^2}{2M}-\frac{i\sigma_2}{M} (\bm{S} \cdot \widehat{\bm{\pi}})^2+\frac{e}{2M^2}(1+\sigma_1) \times \qquad \qquad \qquad \qquad ~~
\nonumber\\
\qquad \qquad \qquad  \left[ iC^{(0)} \left[ \widehat{\bm{E}} \cdot \widehat{\bm{\pi}}-S_i S_j \widehat{\bm{E}}_j \widehat{\bm{\pi}}_i \right]
-2C_1^{(2)}(\overline{\bm{\nabla}} \cdot \widehat{\bm{E}})+2C_3^{(2)}\left[ \overline{\bm{\nabla}} \cdot \widehat{\bm{E}} - \frac{1}{2}(S_i S_j+S_j S_i) \overline{\bm{\nabla}}_i \widehat{\bm{E}}_j \right] \right]
\nonumber\\
 -\frac{ieC^{(0)}}{2M^2}(1-\sigma_1) \left[ \widehat{\bm{\pi}} \cdot \widehat{\bm{E}}-S_i S_j  \widehat{\bm{\pi}}_j\widehat{\bm{E}}_i \right]+
 \mathcal{O}\left(\frac{1}{M^4},F^2\right).~~~
\label{eq:rel-Hamiltonian-E}
\end{eqnarray}
$\widehat{\bm{\pi}}=\widehat{\bm{p}}-eQ_0\widehat{\bm{A}}$ is the conjugate momentum operator corresponding to the spatial covariant derivative, $\bm{D}$, and the $\bm{x}$ coordinate is consequently promoted to a quantum-mechanical operator, $\widehat{\bm{x}}$ (as is any space-dependent function such as the electric field). The $\sigma_i$s are the Pauli matrices and act either on an implicit $3 \times 3$ unity matrix or the spin-1 matrices through a direct multiplication.

The Hamiltonian in Eq. (\ref{eq:rel-Hamiltonian-E}) is comprised of
\begin{eqnarray}
\widehat{\mathcal{H}}_{\text{SR}}=\widehat{\mathcal{E}}^{(-1)}+\widehat{\mathcal{E}}^{(0)}+ \dots+\widehat{\mathcal{O}}^{(1)}+\widehat{\mathcal{O}}^{(2)}+ \dots ,
\label{eq:H-even-odd}
\end{eqnarray}
where $\widehat{\mathcal{E}}^{(n)}$ and $\widehat{\mathcal{O}}^{(n)}$ denote operators that are proportional to $\mathbb{I}_{3\times3},~\sigma_3$ (even) and $\sigma_1,~\sigma_2$ (odd), respectively. The superscript on these operators denote the order at which they contribute in a $\frac{1}{M}$ expansion. The odd operators couple the upper and lower components of the wavefunction in the EOMs. These equations can be decoupled order by order in the $\frac{1}{M}$ expansion using the familiar Foldy-Wouthuysen-Case  (FWC) transformation \cite{Foldy:1949wa, PhysRev.95.1323}). Explicitly, one has
\begin{eqnarray}
\widehat{\mathcal{H}}^{'}={U^{(1)}}^{-1}\widehat{\mathcal{H}}_{\text{SR}}U^{(1)},
\label{eq:H-unitary-trans}
\end{eqnarray}
where the unitary transformation
\begin{eqnarray}
U^{(1)} \equiv e^{i\widehat{\mathcal{S}}^{(1)}} \equiv e^{-\frac{\sigma_3}{2M}\widehat{\mathcal{O}}^{(1)}},
\label{eq:U-unitary-trans}
\end{eqnarray}
removes the odd terms at $\mathcal{O}(1/M)$ in the transformed Hamiltonian, $\widehat{\mathcal{H}}^{'}$, leaving only the odd terms that are of $\mathcal{O}(1/M^{2})$ or higher. The next transformation,
\begin{eqnarray}
U^{(2)} \equiv e^{i\widehat{\mathcal{S}}^{(2)}} \equiv e^{-\frac{\sigma_3}{2M}\widehat{\mathcal{O}}^{(2)}},
\label{eq:U-unitary-trans-II}
\end{eqnarray}
takes the odd operators in $\widehat{\mathcal{H}}^{'}$ and builds a new Hamiltonian, $\widehat{\mathcal{H}}^{''}$, that is free of odd terms also at $\mathcal{O}(1/M^2)$,
\begin{eqnarray}
\widehat{\mathcal{H}}^{''}={U^{(2)}}^{-1}\widehat{\mathcal{H}}^{'}U^{(2)}.
\label{eq:H-unitary-trans-II}
\end{eqnarray}
By iteratively performing this transformation, all the odd operators can be eliminated up to the order one desires. Through this procedure, the NR reduction of the semi-relativistic theory can be systematically obtained.

Following the above procedure, we find that the NR Hamiltonian for the case of a nonzero $\bm{E}$ field up to $\mathcal{O}\left(\frac{1}{M^4}\right)$ is\footnote{A useful formula is the Baker-Campbell-Hausdorff relation,
\begin{eqnarray}
e^{-i\widehat{\mathcal{S}}}\widehat{\mathcal{H}}e^{i\widehat{\mathcal{S}}}=\widehat{\mathcal{H}}-i[ \widehat{\mathcal{S}},\widehat{\mathcal{H}} ]-\frac{1}{2!}[\widehat{\mathcal{S}}, [ \widehat{\mathcal{S}}, \widehat{\mathcal{H}} ] ]+ \dots .
\nonumber
\label{eq:BCH-formula}
\end{eqnarray}
}
\begin{eqnarray}
\widehat{\mathcal{H}}^{(\bm{E})}_{\text{NR}}&=&M \sigma_3+eQ_0\varphi+\sigma_3 \frac{\widehat{\bm{\pi}}^2}{2M}-\sigma_3\frac{\widehat{\bm{\pi}}^4}{8M^3}-\sigma_3\frac{(\bm{S}.\widehat{\bm{\pi}})^4}{2M^3}+\sigma_3\frac{\{ \widehat{\bm{\pi}}^2,(\bm{S} \cdot \widehat{\bm{\pi}})^2 \}}{4M^3}
\nonumber\\
&&-\frac{eC^{(0)}}{4M^2} \left[\bm{S} \cdot (\widehat{\bm{E}} \times \widehat{\bm{\pi}})- \bm{S} \cdot (\widehat{\bm{\pi}} \times \widehat{\bm{E}}) \right]-\frac{e(C^{(0)}+6C_1^{(2)}-2C_3^{(2)})}{6M^2} \overline{\bm{\nabla}} \cdot \widehat{\bm{E}}
\nonumber\\
&&  -\frac{e(-C^{(0)}+2C_3^{(2)})}{4M^2}\left[S_iS_j+S_jS_i-\frac{2}{3} S^2\delta_{ij} \right]\overline{\bm{\nabla}}_i \widehat{\bm{E}}_j+
\mathcal{O}\left(\frac{1}{M^4},F^2\right).
\label{eq:Hamiltonian-nonzero-E}
\end{eqnarray}
Note that, as expected, this Hamiltonian is invariant under parity and time-reversal, and is no longer proportional to  $\sigma_1$ and $\sigma_2$. Additionally, by utilizing the matching conditions in Eq. (\ref{eq:LECs-Rel-I}), (\ref{eq:LECs-Rel-II}) and (\ref{eq:LECs-Rel-IV}), one finds
\begin{eqnarray}
C^{(0)}&=&\overline{\mu}_1-Q_0,
\\
C^{(0)}+6C_1^{(2)}-2C_3^{(2)}&=&\frac{1}{e}M^2\braket{r^2}_E,
\\
-C^{(0)}+2C_3^{(2)}&=&\overline{Q}_2.
\label{eq:LEC-combinations}
\end{eqnarray}
Since the most general effective Lagrangian was used, with low-energy coefficients that are directly matched to the low-energy EM properties of the spin-1 particle, the expected NR interactions are automatically produced with the desired coefficients: the value of $C^{(0)}$ gives the correct coefficient of the spin-orbit interaction in Eq. (\ref{eq:Hamiltonian-nonzero-E}). Moreover, the coefficients of the Darwin term, $\overline{\bm{\nabla}} \cdot \bm{E}$, and the quadrupole interaction, $\left[S_iS_j+S_jS_i-\frac{4}{3} \delta_{ij}\right]\overline{\bm{\nabla}}_i \bm{E}_j$, are correctly produced to be proportional to the particle's mean-squared electric charge radius and the quadrupole moment, respectively.

The coefficient of the Darwin (contact) term we have obtained here differs that obtained by Young and Bludman \cite{Young:1963zza} which is found to be $\frac{1}{6}\overline{Q}_2$ (this reference assumes $\braket{r^2}_E=0$). This is only a definitional issue as  if one defines the electric charge radius in Eq. (\ref{eq:radius-E}) to be the derivative of the $F_1$ form factor with respect to $Q^2$ at $Q^2=0$ (instead of the derivative of the Sachs form factor, $G_C$, that has been adopted here), both results agree.\footnote{We note however that from a physical point of view, these are the Sachs form factors that are directly related to the NR charge and current distributions inside the hadrons, see Eqs. (\ref{eq:rho}) and (\ref{eq:rho-M}), and so the current definitions appear more natural (for a discussion of different definitions and associated confusions see Ref. \cite{Friar:1997js}).} With our definition of the charge radius, the coefficient of the Darwin term for spin-0 and spin-1 particles \cite{Lee:2013lxa} turns out to be the same, both having the value of $-\frac{\braket{r^2}_E}{6}$, which is a convenient feature. After accounting for this difference, the Hamiltonian in Eq. (\ref{eq:Hamiltonian-nonzero-E}) is in complete agreement with those presented in Refs. \cite{Taketani:1940, PhysRev.95.1323, Young:1963zza}, and extends the results in the literature by including all the operators at $\mathcal{O}\left(\frac{1}{M^2}\right)$. The NR Hamiltonian in Eq. (\ref{eq:Hamiltonian-nonzero-E}) applies straightforwardly to scalar particles in an external electric field by setting $\bm{S}=0$.

\subsection{An external magnetic field}
Eqs. (\ref{eq:phi-EOM-E-B}) and (\ref{eq:V-EOM-E-B}) for the case of an external magnetic field that is constant in time can be rewritten as
\begin{eqnarray}
i\frac{d}{dt}\bm{\phi}=M\bm{V}+eQ_0\varphi \bm{\phi}-\frac{1}{M} (\bm{S} \cdot \bm{D})^2 \bm{V}-\frac{eC^{(0)}}{M}(\bm{S} \cdot \bm{B}) \bm{V}
-\frac{2eC_1^{(2)}}{M^3}(\bm{S} \cdot \overline{\bm{\nabla}}) (\bm{B} \cdot \bm{D})\bm{V}
\nonumber\\
 -\frac{eC_2^{(2)}}{M^3} \overline{\bm{\nabla}}^2(\bm{S} \cdot \bm{B}) \bm{V}+
\frac{2eC_3^{(2)}}{M^3}\left[ \overline{\bm{\nabla}}_k (\bm{S} \cdot \bm{B}) \bm{D}_k -\frac{1}{2}(S_i S_j+S_jS_i) \overline{\bm{\nabla}}_j(\bm{S} \cdot \bm{B}) \bm{D}_i \right] \bm{V}
\nonumber\\
-\frac{eC_3^{(2)}}{M^3}\left[ \overline{\bm{\nabla}}_k (\bm{S} \cdot \overline{\bm{\nabla}}) \bm{B}_k -S_i S_j \overline{\bm{\nabla}}_i (\bm{S} \cdot \overline{\bm{\nabla}}) \bm{B}_j \right] \bm{V}+\mathcal{O}\left(\frac{1}{M^4},F^2\right),
\label{eq:EOM-phi-V-nonzero-B-I}
\end{eqnarray}
\begin{eqnarray}
i\frac{d}{dt}\bm{V}=M\bm{\phi}+eQ_0\varphi \bm{V}-\frac{1}{M} \left[ \bm{D}^2-(\bm{S} \cdot \bm{D})^2 \right] \bm{\phi}-\frac{eQ_0}{M}(\bm{S} \cdot \bm{B}) \bm{\phi}+\mathcal{O}\left(\frac{1}{M^4},F^2\right),
\label{eq:EOM-phi-V-nonzero-B-II}
\end{eqnarray}
with the help of spin-1 matrices in Eq. (\ref{eq:spin-matrices}). In terms of the 6-component field $\psi$ introduced in Eq. (\ref{eq:psi-def}), the EOM reads
\begin{eqnarray}
&i\frac{d}{dt}\psi=\widehat{\mathcal{H}}^{(\bm{B})}_{\text{SR}} \psi,
\label{eq:EOM-psi-nonzero-B}
\end{eqnarray}
with the semi-relativistic Hamiltonian
\begin{eqnarray}
&& \widehat{\mathcal{H}}^{(\bm{B})}_{\text{SR}} = M \sigma_3+eQ_0\widehat{\varphi}+(\sigma_3+i\sigma_2)\frac{\widehat{\bm{\pi}}^2}{2M}-\frac{i\sigma_2}{M} (\bm{S} \cdot \widehat{\bm{\pi}})^2
-(\sigma_3-i\sigma_2)\frac{eC^{(0)}}{2M}(\bm{S} \cdot \widehat{\bm{B}})
\nonumber\\
&& \hspace{0.75 cm}  -(\sigma_3+i\sigma_2)\frac{eQ_0}{2M}(\bm{S} \cdot \widehat{\bm{B}})- (\sigma_3-i\sigma_2) \frac{e}{2M^3}\left[2iC_1^{(2)}(\bm{S} \cdot \overline{\bm{\nabla}}) (\widehat{\bm{B}} \cdot \widehat{\bm{\pi}})+C_2^{(2)} \overline{\bm{\nabla}}^2(\bm{S} \cdot \widehat{\bm{B}}) \right .
\nonumber\\
&& \hspace{5 cm} \left .  
-2iC_3^{(2)}[ \overline{\bm{\nabla}}_k (\bm{S} \cdot \widehat{\bm{B}}) \widehat{\bm{\pi}}_k -\frac{1}{2}(S_i S_j+S_jS_i) \overline{\bm{\nabla}}_i (\bm{S} \cdot \widehat{\bm{B}}) \widehat{\bm{\pi}}_j ]+ \right .
\nonumber\\
&& \hspace{4.35 cm} \left . C_3^{(2)}\left[ \overline{\bm{\nabla}}_k (\bm{S} \cdot \overline{\bm{\nabla}}) \widehat{\bm{B}}_k -S_i S_j \overline{\bm{\nabla}}_i (\bm{S} \cdot \overline{\bm{\nabla}}) \widehat{\bm{B}}_j \right] \right]+\mathcal{O}\left(\frac{1}{M^4},F^2\right).
\label{eq:rel-Hamiltonian-B}
\end{eqnarray}

The decoupling of the EOM for the upper and lower three components of $\psi$ can be performed via the FWC procedure as detailed above. The result is
\begin{eqnarray}
&&\widehat{\mathcal{H}}^{(\bm{B})}_{\text{NR}}=M \sigma_3+eQ_0\widehat{\varphi}+\sigma_3 \frac{\widehat{\bm{\pi}}^2}{2M}-\sigma_3\frac{e(C^{(0)}+Q_0)}{2M}(\bm{S} \cdot \widehat{\bm{B}})-\sigma_3\frac{\widehat{\bm{\pi}}^4}{8M^3}-\sigma_3\frac{(\bm{S}.\widehat{\bm{\pi}})^4}{2M^3}+\sigma_3\frac{\{ \widehat{\bm{\pi}}^2,(\bm{S} \cdot \widehat{\bm{\pi}})^2 \}}{4M^3}
\nonumber\\
&& \hspace{4.9 cm} -\sigma_3 \frac{e(C^{(0)}-Q_0)}{8M^3} \left \{ \widehat{\bm{\pi}}^2,\bm{S} \cdot \widehat{\bm{B}} \right \}+
\sigma_3 \frac{e(C^{(0)}-Q_0)}{4M^3}\left \{ (\bm{S} \cdot \widehat{\bm{\pi}})^2,\bm{S} \cdot \widehat{\bm{B}} \right \}+
\nonumber\\
 && \hspace{3.9 cm} -\sigma_3 \frac{e}{2M^3}\left[2iC_1^{(2)}(\bm{S} \cdot \overline{\bm{\nabla}}) (\widehat{\bm{B}} \cdot \widehat{\bm{\pi}})+C_2^{(2)} \overline{\bm{\nabla}}^2(\bm{S} \cdot \widehat{\bm{B}})-2iC_3^{(2)}[ \overline{\bm{\nabla}}_k (\bm{S} \cdot \widehat{\bm{B}}) \widehat{\bm{\pi}}_k  \right .
\nonumber\\
&& \hspace{0.4 cm} \left . -\frac{1}{2}(S_i S_j+S_jS_i) \overline{\bm{\nabla}}_i (\bm{S} \cdot \widehat{\bm{B}}) \widehat{\bm{\pi}}_j ]+C_3^{(2)}\left[ \overline{\bm{\nabla}}_k (\bm{S} \cdot \overline{\bm{\nabla}}) \widehat{\bm{B}}_k -S_i S_j \overline{\bm{\nabla}}_i (\bm{S} \cdot \overline{\bm{\nabla}}) \widehat{\bm{B}}_j \right] \right]+\mathcal{O}\left(\frac{1}{M^4},F^2\right). 
\label{eq:Hamiltonian-nonzero-B}
\end{eqnarray}
Note that the coefficient of the magnetic dipole interaction is correctly produced by recalling that  $C^{(0)}+Q_0=\overline{\mu}_1=\frac{2M}{e}\mu_1$. The coefficients of the rest of the terms are all constrained with the aid of Eqs. (\ref{eq:LECs-Rel-I}-\ref{eq:LECs-Rel-IV}). This Hamiltonian can be reduced to that of spin-0 particles upon setting $\bm{S}=0$.

In contrast with the case of an electric field, the Hamiltonian in a magnetic field depends on all the low-energy coefficients defined in the original relativistic Lagrangian, Eq. (\ref{eq:Lagrangian-Rel-6}), up to this order. For example, it has a dependence on the coefficient $C^{(2)}_2$ at $\mathcal{O}\left(\frac{1}{M^3}\right)$, which according to matching Eq. (\ref{eq:LECs-Rel-III}), is sensitive to the mean-squared magnetic charge radius of the particle, $\braket{r^2}_M$. To obtain this quantity, however, requires introducing a $\bm{B}$ field whose Laplacian is nonzero. In contrast, the dependence on the mean-squared electric charge radius,  $\braket{r^2}_E$, starts at $\mathcal{O}\left(\frac{1}{M^2}\right)$ in the NR Hamiltonian with an $\bm{E}$ field for which only the spatial divergence is required to be nonzero. Although in principle, both the electric and magnetic charge radii could be constrained by generating external fields with proper spatial variations, in the next sections of this paper we focus our interest only on the former. This can be determined at the same order as the electric quadrupole moment through matching to lattice correlation functions in, e.g., a linearly varying $\bm{E}$ field in space, as will be studied in the next section.

\section{Semi-relativistic Green's Functions in Nonuniform External Fields
\label{sec:Rel-Greens-functions} 
}
\noindent
The effective hadronic theory that was set up in the previous sections will be constrained by matching to LQCD correlation functions defined as
\begin{eqnarray}
C_{\alpha\beta}(\bm{x},\tau;\bm{x}',\tau')=\braket{0|[\mathcal{O}_{\psi}(\bm{x},\tau)]_{\alpha}[\mathcal{O}_{\psi^{\dagger}}(\bm{x}',\tau')]_{\beta}|0}_{A_{\mu}}.
\label{eq:correlator-def}
\end{eqnarray}
$[\mathcal{O}_{\psi^{\dagger}}]_{\alpha}$ is an interpolating operator that is constructed from the quark and gluon fields, and creates, out of the vacuum, all states with the same quantum numbers as those of the $\alpha^{\text{th}}$ component of the single-particle state of interest, denoted by $\psi$. Similarly, $[\mathcal{O}_{\psi}]_{\beta}$ acts as a sink that annihilates the $\beta^{\text{th}}$ component of such states. The correlation functions therefore forms $6 \times 6$ matricies according to the above construction of the spin-1 field theory. $A_{\mu}$ denotes the $U(1)$ background gauge field that has been implemented in evaluating the correlation functions, giving rise to background $\bm{E}$ and/or $\bm{B}$ fields, and $\tau$ and $\tau'$ refer to Euclidean times, i.e., $\tau=it$ and $\tau'=it'$.  Assuming that the contributions due to nonvanishing overlap onto states other than the state of interest are small, the correlation function in Eq. (\ref{eq:correlator-def}) directly corresponds to the Green's function of the effective single-particle theory, up to an overall overlap factor. This factor can be cancelled by forming appropriate ratios of correlation functions, see e.g., Ref. \cite{Detmold:2010ts}, and therefore a direct matching of correlation functions of QCD and the Green's functions of the hadronic theory is possible.\footnote{Here we assume that the single-hadron state of interest represents the ground state of the hadronic theory. If the contributions from the excited states are not small, the contribution from the ground state must be isolated or the method of this work will not be applicable. For systems that possess well-defined eigenenergies, this can be achieved by studying the correlation function at large Euclidean times. More complicated analysis is necessary in other cases, in particular when time varying background fields are considered.}

The aim of this section is to construct single-particle Green's functions in the effective theory of the previous sections with a particular external field that gives access to the electric charge radius and the quadrupole moment of the composite spin-1 field. Once the Green's functions are obtained, the next step is to match to correlation functions of a corresponding LQCD calculation. In order to perform such matching, the Green's functions must be first transformed to Euclidean spacetime, $t \rightarrow -i\tau$, and should be modified to correspond to a hadronic system enclosed in a finite (hyper)cubic volume.\footnote{Assuming that the LQCD correlation functions have been already extrapolated to the continuum limit. One can alternatively formulate the hadronic theory away from the continuum limit, see e.g., Ref. \cite{Tiburzi:2012ks}, which, in general, gives rise to more complexities, partly due to new interactions that must be introduced.} Furthermore, since PBCs are commonly imposed on the fields in LQCD calculations, the Green's functions of the hadronic theory should be constructed in such a way to satisfy these boundary conditions.\footnote{In general, this requires a special treatment of the $U(1)$ background gauge links near the boundary of the lattice, as well as implementing appropriate quantization conditions on the parameters of the background fields, to guarantee the full periodicity of the correlation functions, see Ref. \cite{Davoudi:2015cba}.} In general,  solving the hadronic theory in a finite volume is more involved and so one hopes that the knowledge of the infinite-volume Green's functions is sufficient to form the FV counterparts. Indeed, when the particle's wavefunction is localized within the volume in a given external potential (so that the Green's functions are suppressed at the boundary of the volume),\footnote{An explicit example will be provided in the next section for systems that do not exhibit such a feature.} the FV Green's functions can be seen to simply arise by forming a sum over the periodic images of the infinite-volume Green's functions. Explicitly,
\begin{eqnarray}
G^{V}(\bm{x},\tau;\bm{x}',\tau') = \sum_{\nu,\bm{n}}G(\bm{x}+\bm{n}L,\tau+\nu T;\bm{x}',\tau';M^{V},Q_2^V,\braket{r^2}_E^V, \dots),
\label{eq:rel-FV-GF-images}
\end{eqnarray}
so that 
\begin{eqnarray}
G^{V}(\bm{x}+\bm{n}L,\tau+\nu T;\bm{x}',\tau')=G^{V}(\bm{x},\tau;\bm{x}',\tau').
\label{eq:rel-FV-GF-BCs}
\end{eqnarray}
$\nu$ is an integer, $\bm{n}$ is a triplet of integers, and $T$ and $L$ denote the temporal and spatial extents of the volume, respectively. Note that the mass, as well as the low-energy EM couplings that enter the FV Green's functions differ from those in an infinite volume. As a result, $G(\bm{x},\tau;\bm{x}',\tau';M^{V},Q_2^V,\braket{r^2}_E^V, \dots)$ is the infinite-volume Green's function that is evaluated at the FV values for the mass, the quadrupole moment,  the mean-squared charge radius, etc., as indicated by the superscript $V$ on these quantities. Unfortunately, while the general single-particle hadronic theory is useful in analyzing the correlation functions, it is of no help in identifying the FV corrections to these couplings, and a more elaborate EFT is required for that purpose. These effective theories describe the substructure of hadrons and nuclei that have been neglected in the single-particle description here, and are therefore specific to each particle. In the case of uniform background fields, for example, these corrections have been evaluated for pions with the use of chiral perturbation theory, and the required infrared renormalization of charge and polarizabilities are obtained in a finite volume \cite{Detmold:2006vu, Tiburzi:2008pa}. These volume corrections are exponentially suppressed for the case of stable particles. For single hadrons, the leading exponential suppression is governed by the pion mass, while for the case of bound states such as the deuteron, they are governed also by the state's binding momentum. When applying the method of this paper to each physical system, one must quantify these corrections with the aid of the appropriate low-energy theory, or alternatively perform calculations at multiple lattice volumes to allow for numerical extrapolations. Nonetheless, as long as the volume is large compared with the aforementioned scales, the EM couplings of the single hadron theory, as extracted from matching the FV Green's function to LQCD correlation functions, can be approximated by their infinite-volume values.

The goal of this section is to derive the functional form of the semi-relativistic Green's functions in an infinite-volume, $G(\bm{x},\tau;\bm{x}',\tau';M,Q_2,\braket{r^2}_E, \dots)$, but no attempt will be made to quantify the volume-dependence of the mass and low-energy parameters here. Although the connection between the semi-relativistic Green's functions and LQCD correlation functions is clear, their form, even in the simplest nonuniform fields considered, is complicated, as will be seen in this section, making the matching procedure somewhat nontrivial. On the other hand, the NR Green's functions must be related to LQCD correlation functions indirectly, but likely lead to more straightforward matching procedures (assuming  the system is NR to a good approximation).

\subsection{A spin-1 field in the absence of external fields
}
To begin with, let us focus on the noninteracting case and obtain the relativistic Green's function for the spin-1 theory that was formulated in the previous section using a 6-component field, $\psi$. The degrees of freedom of the field $\psi$ correspond to a relativistic spin-1 field. As a result, one should find that despite a first-order EOM, Eq. (\ref{eq:EOM-psi-nonzero-E}), each component of the field indeed acquires a relativistic dispersion relation. After setting $\bm{E}=0$ in Eq. (\ref{eq:rel-Hamiltonian-E}), the EOM for the field $\psi$ reduces to
\begin{eqnarray}
i\frac{d}{dt}\psi(\bm{x},t)= \left \{ M \sigma_3-(\sigma_3+i\sigma_2)\frac{\bm{\nabla}^2}{2M}+\frac{i\sigma_2}{M} (\bm{S} \cdot \bm{\nabla})^2 \right \} \psi(\bm{x},t),
\label{eq:EOM-psi-E-zero}
\end{eqnarray}
with a corresponding EOM for the relativistic Green's function
\begin{eqnarray}
\left \{ i\frac{d}{dt}-M \sigma_3+(\sigma_3+i\sigma_2)\frac{\bm{\nabla}^2}{2M}-\frac{i\sigma_2}{M} (\bm{S} \cdot \bm{\nabla})^2 \right \} G(\bm{x},t;\bm{x}',t')=i\delta(\bm{x}-\bm{x}')\delta(t-t').
\label{eq:EOM-GF-E-zero}
\end{eqnarray}

Following the method by Schwinger \cite{Schwinger:1951nm}, one can regard $G(\bm{x},t;\bm{x}',t')$ as the matrix element of an operator $\widehat{G}$ that acts on states labeled by spacetime coordinates, as well as vectorial indices which will be suppressed in the following. Explicitly,
\begin{eqnarray}
G(\bm{x},t;\bm{x}',t') \equiv \braket{\bm{x},t|\widehat{G}|\bm{x}',t'},
\end{eqnarray}
where the operator $\widehat{G}$ is simply
\begin{eqnarray}
\widehat{G}&=&\frac{i}{\widehat{p}_0-M \sigma_3-(\sigma_3+i\sigma_2)\frac{\widehat{\bm{p}}^2}{2M}+\frac{i\sigma_2}{M} (\bm{S} \cdot \widehat{\bm{p}})^2},
\end{eqnarray}
with $\widehat{p}_0 \equiv i \frac{d}{dt} $ and $\widehat{\bm{p}} \equiv -i\bm{\nabla}$, where $\bm{\nabla}$ is the gradient operator with respect to the $\bm{x}$ coordinate. It is now straightforward to show that the operator $\widehat{G}$ can be expressed in the following form
\begin{eqnarray}
\widehat{G}&=&\left[\widehat{p}_0+M \sigma_3+(\sigma_3+i\sigma_2)\frac{\widehat{\bm{p}}^2}{2M}-\frac{i\sigma_2}{M} (\bm{S} \cdot \widehat{\bm{p}})^2\right]\frac{i}{\widehat{p}_0^2-\widehat{\bm{p}}^2-M^2},
\end{eqnarray}
from which it is manifest that each mode of the field $\psi$ satisfies a relativistic dispersion relation, $p_0^2=\bm{p}^2+M^2$. Moreover, in this form it is evident that the Green's function of the spin-1 field can be simply deduced from that of the spin-0 field. The coordinate-space Green's function can be obtained using the familiar Schwinger formalism, where one devolves the evaluation of $G(\bm{x},t;\bm{x}',t')$ to a corresponding quantum-mechanical system evolving in a proper time, $s$, with a proper \emph{Hamiltonian}, 
\begin{eqnarray}
G(\bm{x},t;\bm{x}',t') &=& \braket{\bm{x},t|\left[\widehat{p}_0+M \sigma_3+(\sigma_3+i\sigma_2)\frac{\widehat{\bm{p}}^2}{2M}-\frac{i\sigma_2}{M} (\bm{S} \cdot \widehat{\bm{p}})^2\right]
\int_0^{\infty} ds ~ e^{i(\widehat{p}_0^2-\widehat{\bm{p}}^2-M^2+i\epsilon)s}
|\bm{x}',t'}
\nonumber\\
 &=& \left [ i\frac{d}{dt}+M \sigma_3-(\sigma_3+i\sigma_2)\frac{\bm{\nabla}^2}{2M}+\frac{i\sigma_2}{M} (\bm{S} \cdot \bm{\nabla})^2 \right ] \int_0^{\infty} ds ~ e^{-i(M^2-i\epsilon)s} \braket{\bm{x},t|e^{-i\widehat{\mathcal{H}}^{(0)}s}|\bm{x}',t'}.
 \nonumber\\
\end{eqnarray}
The proper-time Hamiltonian for this noninteracting case simply is $\widehat{\mathcal{H}}^{(0)} \equiv \frac{-\widehat{p}_0^2+\widehat{\bm{p}}^2}{2m}$ with a corresponding mass $m=\frac{1}{2}$. Therefore,
\begin{eqnarray}
\braket{\bm{x},t|e^{-i\widehat{\mathcal{H}}^{(0)}s}|\bm{x}',t'} \equiv \braket{\bm{x},t;s|\bm{x}',t';0} ,
\end{eqnarray}
is nothing but the well-known quantum-mechanical propagator of a free particle with mass $m$ in four spacetime dimensions. Consequently, the Green's function evaluates to
\begin{eqnarray}
G(\bm{x},t;\bm{x}',t') &=& i \left [ i\frac{d}{dt}+M \sigma_3-(\sigma_3+i\sigma_2)\frac{\bm{\nabla}^2}{2M}+\frac{i\sigma_2}{M} (\bm{S} \cdot \bm{\nabla})^2 \right ]
\nonumber\\
&& \hspace{1.5 cm} \times \int_0^{\infty} ds ~ e^{-i(M^2-i\epsilon)s} \left(\frac{1}{4 \pi i s}\right)^2 e^{\frac{(t-t')^2-(\bm{x}-\bm{x}')^2}{4is}}.
\label{eq:G-free-Minkowski}
\end{eqnarray}
%

\subsection{A charged spin-1 field coupled to a linearly varying electric field in space
}
Recalling that our goal is to constrain the charge radius and the electric quadrupole moment of the composite spin-1 particle by matching the semi-relativistic Green's functions to lattice correlation functions,  we now turn to the case of a composite spin-1 field immersed in an external electric field. As is evident from Eq. (\ref{eq:rel-Hamiltonian-E}), in order to access these quantities, it suffices to consider an electric field that has only a nonzero spatial gradient. A simple choice for the background EM gauge potential is 
\begin{eqnarray}
\varphi=-\frac{1}{2}E_0\bm{x}_3^2,~\bm{A}=\bm{0},
\label{eq:A-field-NU}
\end{eqnarray}
so that a linearly varying field in space is generated along the $\bm{x}_3$ direction,
\begin{eqnarray}
\bm{E}=E_0\bm{x}_3\hat{\bm{x}_3},
\label{eq:E-field-NU}
\end{eqnarray}
providing a constant field gradient, $E_0$.\footnote{With this choice of the gauge field, no Wick rotation of the electric field is needed when one transforms to Euclidean spacetime. Note that the boldfaced quantities denote  three-vectors. When they are assigned a subscript, they refer to the \emph{components} of a three-vector. This convention is used throughout to distinguish the components of a three-vector form those of a Minkowski four-vector. Explicitly, $\bm{x}_3=-x_3$, and so on.} As already mentioned, such a space-dependent field creates a large electric field strength at large values of $\bm{x}_3$, causing effects whose description is beyond the effective single-particle theory of this paper. In the following discussions, we restrict ourselves to a finite region of spacetime such that as long as  the field slope $E_0$ is tuned properly, the field strength remains weak compared with the characteristic scales of the low-energy theory. The situation is further controlled in a finite volume where the field slop can be set to scale with inverse powers of volume such that no strong electric field is produced anywhere within the volume.

With the choice of the gauge potential in Eq. (\ref{eq:A-field-NU}), the EOM of field $\bm{\psi}$ does not depend on the $\bm{x}_1$ and $\bm{x}_2$ coordinates. Accordingly, the conjugate momenta corresponding to these coordinates, namely $\bm{p}_1$ and $\bm{p}_2$, are conserved quantum numbers of the system, and the wavefunctions are simply plane waves along these directions. Consequently, for the Green's function it follows that
\begin{eqnarray}
G(\bm{x},t;\bm{x}',t')=\frac{1}{(2\pi)^2}
\int d\bm{p}_1 d\bm{p}_2~
e^{i\bm{p}_1(\bm{x}_1-\bm{x}_1')+i\bm{p}_2(\bm{x}_2-\bm{x}'_2)}\widetilde{G}(\bm{x}_3,t;\bm{x}_3',t';\bm{p}_1,\bm{p}_2).
\label{eq:GF-plane-waves-rel}
\end{eqnarray}
The dependence of $\widetilde{G}$ on transverse momenta  $\bm{p}_1$ and $\bm{p}_2$ can in general be nontrivial, however one can choose to project onto the sector of zero transverse momentum, $\bm{p}_1=\bm{p}_2=0$. This enables one to take advantage of substantial simplifications that arise without losing sensitivity to the charge radius and quadrupole moment interactions as will be seen below. The desired projection can be obtained by integrating both sides of Eq. (\ref{eq:GF-plane-waves-rel}) over $\bm{x}_1$ and $\bm{x}_2$. The EOM for the projected Green's function $\widetilde{G}(\bm{x}_3,t;\bm{x}_3',t';\bm{p}_1,\bm{p}_2)$ in the zero transverse-momentum sector can be readily deduced from the EOM of the field in Eq. (\ref{eq:EOM-psi-nonzero-E}). Explicitly,
\begin{align}
& \left \{ i\frac{d}{dt}-M \sigma_3+\frac{1}{2}M\omega_E^2\bm{x}_3^2+(\sigma_3+i\sigma_2)\frac{1}{2M}\frac{d^2}{d\bm{x}_3^2}-i\sigma_2\frac{1}{M}\frac{d^2}{d\bm{x}_3^2}S_3^2 \right.
\nonumber\\
& \left . -(1+\sigma_1)\frac{e}{2M^2} \left[ C^{(0)} E_0\bm{x}_3 \frac{d}{d\bm{x}_3}(1-S_3^2) 
-2C_1^{(2)}E_0+2C_3^{(2)}E_0(1-S_3^2) \right] + \right.
\nonumber\\
& \left . (1-\sigma_1)\frac{eC^{(0)}}{2M^2}E_0\bm{x}_3 \frac{d}{d\bm{x}_3}(1-S_3^2) \right \} \widetilde{G}(\bm{x}_3,t;\bm{x}_3',t';\bm{p}_1=\bm{p}_2=0)=i\delta(\bm{x}_3-\bm{x}_3')\delta(t-t'),
\label{eq:EOM-GF-E-x2}
\end{align}
where we have defined
\begin{eqnarray}
\omega_E^2 \equiv \frac{eQ_0E_0}{M}.
\label{eq:omega-E-def}
\end{eqnarray}

As was done in the noninteracting case, one can define an operator $\widehat{\widetilde{G}}$ that acts on the states labeled by the $(\bm{x}_3,t)$ coordinates, and by vectorial indices which will be suppressed in the following. Explicitly,
\begin{eqnarray}
\widetilde{G}(\bm{x}_3,t;\bm{x}'_3,t';\bm{p}_1=\bm{p}_2=0) \equiv \braket{\bm{x}_3,t|\widehat{\widetilde{G}}|\bm{x}'_3,t'},
\end{eqnarray}
where operator $\widehat{\widetilde{G}}$ can be deduced from Eq. (\ref{eq:EOM-GF-E-x2}),
\begin{eqnarray}
\widehat{\widetilde{G}}
&\equiv & \frac{i}{\widehat{p}_0-A^{(0)}(\widehat{\bm{x}}_3)-A^{(2)}-\sigma_1B(\widehat{\bm{x}}_3,\widehat{\bm{p}}_3)-i\sigma_2C(\widehat{\bm{p}}_3)-\sigma_3D(\widehat{\bm{p}}_3)}
\nonumber\\
&=& \left[ 1+i(A^{(2)}+\sigma_1B(\widehat{\bm{x}}_3,\widehat{\bm{p}}_3)) \widehat{\widetilde{G}}^{(0)}+ \dots \right]\widehat{\widetilde{G}}^{(0)}.
\label{eq:GF-E-x2-expand}
\end{eqnarray}
We have defined
\begin{eqnarray}
\label{eq:A0-C-D-def}
&& A^{(0)}(\widehat{\bm{x}}_3) \equiv -\frac{1}{2}M\omega_E^2\widehat{\bm{x}}_3^2,~~~~C(\widehat{\bm{p}}_3) \equiv (1-2S_3^2)\frac{\widehat{\bm{p}}_3^2}{2M},~~~~D(\widehat{\bm{p}}_3) \equiv  M+\frac{\widehat{\bm{p}}_3^2}{2M},
\\
\label{eq:A2-def}
&& A^{(2)} \equiv \frac{e}{2M^2}\left[ -2C_1^{(2)}E_0+2C_3^{(2)}E_0(1-S_3^2) \right],~
\\
\label{eq:B-def}
&& B(\widehat{\bm{x}}_3,\widehat{\bm{p}}_3) \equiv \frac{e}{2M^2}\left[ 2iC^{(0)} (1-S_3^2) E_0\widehat{\bm{x}}_3 \widehat{\bm{p}}_3-
2C_1^{(2)}E_0+2C_3^{(2)}E_0(1-S_3^2) \right],
\end{eqnarray}
and $\widehat{\widetilde{G}}^{(0)}$ in the second line of Eq. (\ref{eq:GF-E-x2-expand}) denotes the projected Green's function of a \emph{structureless} spin-1 particle in this particular external field,
\begin{eqnarray}
\widehat{\widetilde{G}}^{(0)} & \equiv & \frac{i}{\widehat{p}_0-A^{(0)}(\widehat{\bm{x}}_3)-i\sigma_2C(\widehat{\bm{p}}_3)-\sigma_3D(\widehat{\bm{p}}_3)}.
\label{eq:GF-0-rel-I}
\end{eqnarray}
The ellipses in Eq. (\ref{eq:GF-E-x2-expand}) denote terms of $\mathcal{O}\left(\frac{E_0^2}{M^4} \right)$ or higher. $A^{(2)}$ and $B$ terms, being of $\mathcal{O}\left(\frac{E_0}{M^2} \right)$, are suppressed compared with $A^{(0)}$, $C$ and $D$ terms in $\widehat{\widetilde{G}}$ inverse, as long as the electric-field slope is sufficiently small. It is worth noting that a periodic implementation of this background field in LQCD calculations requires a quantization condition to be placed on the electric-field slope, $E_0$. This quantization condition constrains the smallest quantum of this parameter to be proportional to $\frac{1}{L^2T}$, see Ref. \cite{Davoudi:2015cba}. For this scenario, terms of $\mathcal{O}\left(\frac{E_0}{M^2} \right)$ are suppressed by $\sim \frac{1}{M^3L^2T}$ compared with the leading term in $\widehat{\widetilde{G}}$ inverse, and by $\sim \frac{1}{MT}$ compared with the next-to-leading order terms, as long as $L,T \gg \frac{1}{M}$. In this limit, neglecting higher order terms in the expansion in Eq. (\ref{eq:GF-E-x2-expand}) can be regarded to be a reasonable approximation, and is consistent with the organization of nonminimal interactions  in the semi-relativistic Lagrangian as pursued in the previous sections. It must be noted that the electric-field dependent term $\frac{1}{2}M\omega_E^2\widehat{\bm{x}}_3^2$ in $A^{(0)}$ can not be made arbitrarily small compared with the leading terms when a periodic implementation of the electric field in a finite volume is pursued. In fact, this term while being suppressed by $\sim \frac{1}{MT}$ compared with the leading order term in the $MT \gg 1$ limit, turns out to be of the same order as the next-to-leading order contributions. Consequently, such term can not be treated as a small perturbation to the noninteracting Hamiltonian of the system. As we will see below, this term is the source of a nonperturbative quartic potential in the corresponding proper-time quantum-mechanical system.

In order to evaluate $\widehat{\widetilde{G}}^{(0)}$, one may first note that Eq. (\ref{eq:GF-0-rel-I}) can be rewritten as
\begin{eqnarray}
\widehat{\widetilde{G}}^{(0)} 
&=&\frac{i(\widehat{p}_0-A^{(0)}+i\sigma_2C+\sigma_3D)}
{(\widehat{p}_0-A^{(0)})^2+C^2-D^2-[A^{(0)},i\sigma_2C]-[A^{(0)},\sigma_3D]
-\{i\sigma_2C,\sigma_3D\}},
\label{eq:GF-0-rel-II}
\end{eqnarray}
where the commutators readily evaluate to
\begin{align}
& [A^{(0)},i\sigma_2C]  =  
-i\sigma_2(1-2S_3^2)\frac{\omega_E^2}{2}(1+2i \widehat{\bm{x}}_3 \widehat{\bm{p}}_3),
\\
&[A^{(0)},\sigma_3D]  =   
-\sigma_3\frac{\omega_E^2}{2}(1+2i \widehat{\bm{x}}_3 \widehat{\bm{p}}_3),
\end{align}
while the anticommutator evaluates to zero. Substituting these relations in Eq. (\ref{eq:GF-0-rel-II}) and performing further manipulations give
\begin{eqnarray}
&& \widehat{\widetilde{G}}^{(0)} 
= \frac{1}{6}
\left[ \left((\widehat{p}_0+\frac{1}{2}M\omega_E^2\widehat{\bm{x}}_3^2)^2-M^2-\widehat{\bm{p}}_3^2\right)^2+2i\sigma_2(1-2S_3^2)\omega_E^2\widehat{\bm{p}}_3^2+2\sigma_3\omega_E^2\widehat{\bm{p}}_3^2 \right]
\nonumber\\
&& \hspace{1.05 cm} \times ~ \left[(\widehat{p}_0+\frac{1}{2}M\omega_E^2\widehat{\bm{x}}_3^2)^2-M^2-\widehat{\bm{p}}_3^2-i\sigma_2(1-2S_3^2)\frac{\omega_E^2}{2}(1+2i \widehat{\bm{x}}_3 \widehat{\bm{p}}_3)-\sigma_3\frac{\omega_E^2}{2}(1+2i \widehat{\bm{x}}_3 \widehat{\bm{p}}_3)\right]
\nonumber\\
&& \hspace{1.05 cm}  \times ~
\left[\widehat{p}_0+\frac{1}{2}M\omega_E^2\widehat{\bm{x}}_3^2+i\sigma_2 (1-2S_3^2)\frac{\widehat{\bm{p}}_3^2}{2M}+\sigma_3 \frac{\widehat{\bm{p}}_3^2}{2M}\right] \left(\frac{d}{dM^2}\right)^3\widehat{\widetilde{G}}^{(0)}_{scl},
\label{eq:GF-0-rel-III}
\end{eqnarray}
where $\widehat{\widetilde{G}}^{(0)}_{scl}$ is the projected Green's function of a structureless \emph{spin-0} particle in this external field
\begin{eqnarray}
\widehat{\widetilde{G}}^{(0)}_{scl} & \equiv & \frac{i}{(\widehat{p}_0+\frac{1}{2}M\omega_E^2\widehat{\bm{x}}_3^2)^2-M^2-\widehat{\bm{p}}_3^2}.
 \label{eq:GF-0-rel-scalar-I}
\end{eqnarray}
Thus, the problem of finding the Green's function of a composite spin-1 field in the (weak) external electric field of Eq. (\ref{eq:E-field-NU}) has reduced to that of finding $\widehat{\widetilde{G}}^{(0)}_{scl}$. Note that the term $\frac{1}{2}M\omega_E^2\bm{x}_3^2$ in the denominator is independent of $M$, see Eq. (\ref{eq:omega-E-def}), and so the derivative with respect to $M^2$ only acts on the $M^2$ term in $(\widehat{\widetilde{G}}^{(0)}_{scl})^{-1}$.

The projected Green's function of the structureless spin-0 particle in coordinate space can now be obtained from a corresponding quantum-mechanical system that evolves in the proper time, $s$. Explicitly,
\begin{eqnarray}
\widetilde{G}^{(0)}_{scl}(\bm{x}_3,t;\bm{x}'_3,t';\bm{p}_1=\bm{p}_2=0) &=& \braket{\bm{x}_3,t|
\int_0^{\infty} ds ~ e^{i\left((\widehat{p}_0+\frac{1}{2}M\omega_E^2\bm{x}_3^2)^2-M^2-\widehat{\bm{p}}_3^2+i\epsilon \right)s}|\bm{x}'_3,t'}
\nonumber\\
&=& \int_0^{\infty} ds~ e^{-i(M^2-i\epsilon)s} \int_{-\infty}^{\infty} \frac{dp_0}{2\pi}~ e^{-ip_0(t-t')+ip_0^2s} \braket{\bm{x}_3|e^{-i\widehat{\mathcal{H}}^{(\bm{E})}s}|\bm{x}'_3},
\label{eq:GF-0-rel-scalar-II}
\end{eqnarray}
where $\widehat{\mathcal{H}}^{(\bm{E})}=\frac{\widehat{\bm{p}}_3^2}{2m}+\frac{1}{2}m\Omega^2 \widehat{\bm{x}}_3^2+\lambda \widehat{\bm{x}}_3^4$ denotes the quantum-mechanical Hamiltonian of a one-dimensional anharmonic oscillator with mass $m=\frac{1}{2}$, harmonic frequency $\Omega^2=-4p_0M\omega_E^2$ and the quartic coupling $\lambda = -\frac{1}{4}M^2\omega_E^4$. Therefore, once the quantum-mechanical propagator of the anharmonic oscillator, i.e., $\braket{\bm{x}_3|e^{-i\widehat{\mathcal{H}}^{(\bm{E})}s}|\bm{x}'_3}$, is known for all values of $\bm{x}_3$, $\bm{x}'_3$ and $s$, the Green's functions of the spin-1 field in this EM potential can, in principle, be obtained from Eqs. (\ref{eq:GF-0-rel-scalar-II}), (\ref{eq:GF-0-rel-III}) and (\ref{eq:GF-E-x2-expand}). Explicitly,
\begin{eqnarray}
&&\widetilde{G}(\bm{x}_3,t;\bm{x}'_3,t';\bm{p}_1=\bm{p}_2=0)
=\widetilde{G}^{(0)}(\bm{x}_3,t;\bm{x}'_3,t';\bm{p}_1=\bm{p}_2=0)+i\left[A^{(2)}+\sigma_1B(\bm{x}_3,-i\frac{d}{d\bm{x}_3})\right]
\nonumber\\
&& \qquad \qquad \qquad ~~~ \times \int dt'' d\bm{x}_3''~ \widetilde{G}^{(0)}(\bm{x}_3,t;\bm{x}''_3,t'';\bm{p}_1=\bm{p}_2=0)\widetilde{G}^{(0)}(\bm{x}''_3,t'';\bm{x}'_3,t';\bm{p}_1=\bm{p}_2=0),
\label{eq:GF-E-x2-complete-I}
\end{eqnarray}
where
\begin{eqnarray}
&& \widetilde{G}^{(0)}(\bm{x}_3,t;\bm{x}'_3,t';\bm{p}_1=\bm{p}_2=0) 
= \frac{1}{6}
\left[ \left((i\frac{d}{dt}+\frac{1}{2}M\omega_E^2\widehat{\bm{x}}_3^2)^2+M^2\frac{d^2}{d\bm{x}_3^2}\right)^2-2i\sigma_2(1-2S_3^2)\omega_E^2\frac{d^2}{d\bm{x}_3^2} \right .
\nonumber\\
&& \hspace{2.5 cm} \left. -2\sigma_3\omega_E^2\frac{d^2}{d\bm{x}_3^2} \right] 
\times \left[(i\frac{d}{dt}+\frac{1}{2}M\omega_E^2\bm{x}_3^2)^2-M^2+\frac{d^2}{d\bm{x}_3^2}-i\sigma_2\frac{\omega_E^2}{2}(1-2S_3^2)-\sigma_3\frac{\omega_E^2}{2}\right]
\nonumber\\
&& \hspace{0.25 cm} \times \left[i\frac{d}{dt}+\frac{1}{2}M\omega_E^2\bm{x}_3^2-\frac{i\sigma_2}{2M}\frac{d^2}{d\bm{x}_3^2}(1-2S_3^2)-\frac{\sigma_3}{2M}\frac{d^2}{d\bm{x}_3^2}\right]
\left(\frac{d}{dM^2}\right)^3\widetilde{G}^{(0)}_{scl}(\bm{x}_3,t;\bm{x}'_3,t';\bm{p}_1=\bm{p}_2=0),
\label{eq:GF-E-x2-complete-II}
\end{eqnarray}
with $\widetilde{G}^{(0)}_{scl}$ given in Eq. (\ref{eq:GF-0-rel-scalar-II}). Nonetheless, as is well known, there is no closed analytic form for the propagator in an anharmonic oscillator potential (except in the semi-classical limit). Although numerical solutions are plausible and have been studied extensively in literature, the lack of an analytic solution can hinder the evaluation of the Green's function in Eq. (\ref{eq:GF-E-x2-complete-I}) in practice. This is because the propagator must be integrated over the $p_0$ momenta as well as the proper time, $s$, see Eq. (\ref{eq:GF-0-rel-scalar-II}), and should be acted subsequently by the spatial derivative operators, see Eqs. (\ref{eq:GF-E-x2-complete-I}) and (\ref{eq:GF-E-x2-complete-II}). Moreover, to access the contributions that are sensitive to the structure of the spin-1 field, further integrations over the intermediate position and time coordinates are required according to Eq. (\ref{eq:GF-E-x2-complete-I}). Therefore, as far as the matching to LQCD correlation functions is concerned, it is useful to look for alternative approaches that are simpler to be implemented in practice. This will be pursued in Sec. \ref{sec:NR-Greens-functions}.

\subsection{A neutral spin-1 field coupled to a linearly varying electric field in space
}
In contrast to the case of a charged spin-1 field, the semi-relativistic Green's function of a neutral spin-1 field has a simpler form in the electric field considered in Eq. (\ref{eq:E-field-NU}). This Green's function can be deduced from equations above by setting $\omega_{E}=0$. Explicitly,
\begin{eqnarray}
&&\widetilde{G}_{neut}(\bm{x}_3,t;\bm{x}'_3,t';\bm{p}_1=\bm{p}_2=0)
=\widetilde{G}^{(0)}(\bm{x}_3,t;\bm{x}'_3,t';\bm{p}_1=\bm{p}_2=0;\omega_E=0)+i\left[A^{(2)}+\sigma_1B(\bm{x}_3,-i\frac{d}{d\bm{x}_3})\right]
\nonumber\\
&& \qquad ~~~ \times \int dt'' d\bm{x}_3''~ \widetilde{G}^{(0)}(\bm{x}_3,t;\bm{x}''_3,t'';\bm{p}_1=\bm{p}_2=0;\omega_E=0)\widetilde{G}^{(0)}(\bm{x}''_3,t'';\bm{x}'_3,t';\bm{p}_1=\bm{p}_2=0;\omega_E=0),
\end{eqnarray}
where
\begin{eqnarray}
&& \widetilde{G}^{(0)}(\bm{x}_3,t;\bm{x}'_3,t';\bm{p}_1=\bm{p}_2=0;\omega_E=0) 
= \left[i\frac{d}{dt}-\frac{i\sigma_2}{2M}\frac{d^2}{d\bm{x}_3^2}(1-2S_3^2)-\frac{\sigma_3}{2M}\frac{d^2}{d\bm{x}_3^2}\right]
\nonumber\\
&& \qquad \qquad \qquad \qquad \qquad \qquad \qquad \qquad \qquad \qquad \qquad \times ~ \widetilde{G}^{(0)}_{scl}(\bm{x}_3,t;\bm{x}'_3,t';\bm{p}_1=\bm{p}_2=0;\omega_E=0),
\end{eqnarray}
and the projected coordinate-space Green's function of a neutral structureless spin-0 particle simply evaluates to
\begin{eqnarray}
\widetilde{G}^{(0)}_{scl}(\bm{x}_3,t;\bm{x}'_3,t';\bm{p}_1=\bm{p}_2=0;\omega_{E}=0) &=& i \int_0^{\infty} ds ~  e^{-i(M^2-i\epsilon)s} \left(\frac{1}{4 \pi i s}\right) e^{\frac{(t-t')^2-(\bm{x}_3-\bm{x}_3')^2}{4is}}.
\label{eq:GF-0-rel-scalar-III}
\end{eqnarray}

\
\

A feature of the Green's functions in external fields is that while being gauge variant, their gauge dependency can be identified and separated from their gauge-independent part as a phase factor. Further, these Green's functions are not translationally invariant in the presence of spatially nonuniform external fields. We devote Appendix \ref{App:Gauge-dependency} of this paper to make these features more apparent.

\section{Nonrelativistic Green's Functions in Nonuniform External Fields
\label{sec:NR-Greens-functions} 
}
\noindent
Given the challenge associated with directly matching LQCD correlation functions to the semi-relativistic Green's function of the hadronic theory with the chosen nonuniform field of the previous section (in particular for the case of charged spin-1 fields), it may be useful to consider other alternatives. One such alternative, that is only applicable to NR systems, is to consider the Green's functions of the single-particle hadronic theory obtained from the NR Hamiltonians of Sec. \ref{sec:EOM}. However, given the relativistic nature of LQCD calculations, the connection between these Green's functions and LQCD correlation functions must be determined. This section is devoted to such investigations, and aims to identify optimal strategies that lead to constraining the quadrupole moment and the electric charge radius of the composite spin-1 particle from LQCD calculations in background fields.

Let us separate the EOM in the $\bm{E}$ field, Eq. (\ref{eq:Hamiltonian-nonzero-E}), for the upper and lower three components of wavefunctions. In the limit of no external field, there will be two sets of solutions corresponding to positive and negative energy eigenvalues, i.e., $\mathcal{E}^{(\pm)} \sim \pm M$ in the NR limit. When a weak external field is introduced, the first term in Eq. (\ref{eq:Hamiltonian-nonzero-E}) dominates and such distinction still holds.\footnote{The system may not possess eigenenergies as will be seen shortly; making such distinction ambiguous. However, we continue to use the positive- and negative-energy notation for the solutions of the EOM in such cases as well, as motivated by the behavior of solutions in the zero external field limit.} The weak-field assumption for the case of a linearly varying field in $\bm{x}_3$ does not obviously hold as $\bm{x}_3 \rightarrow \infty$, unless the adiabatic procedure of Sec. \ref{subsec:semi-rel-EFT} is used to introduce fields at infinity. Alternatively, as discussed above, the formalism presented here may be restricted to a finite region of space where the strength of the field remains small compared with the square of the mass and the compositeness scale of the particle. Denoting the upper-component wavefunction  with positive-energy eigenvalues by $\bm{\psi}^{(+)}$, and the lower-component wavefunction  with negative-energy eigenvalues by $\bm{\psi}^{(-)}$, one obtains
\begin{eqnarray}
i\frac{d}{dt}\bm{\psi}^{(\pm)}_{\text{NR}}=\pm\widehat{\mathcal{H}}_{\text{NR}}^{(\pm)}\bm{\psi}^{(\pm)}_{\text{NR}},
\label{eq:psi-EOM-pm}
\end{eqnarray}
where $\bm{\psi}_{\text{NR}}^{(\pm)}$ are now three-component wavefunctions, and $\widehat{\mathcal{H}}_{\text{NR}}^{(\pm)}$ are $3 \times 3$ matrices which can be read from the NR Hamiltonian in Eq. (\ref{eq:Hamiltonian-nonzero-E}),
\begin{eqnarray}
\widehat{\mathcal{H}}_{\text{NR}}^{(\pm)}&=& M ~ \mathbb{I}_{3\times3} \pm eQ_0\varphi~\mathbb{I}_{3\times3}+\frac{\widehat{\bm{\pi}}^2}{2M}~\mathbb{I}_{3\times3} \mp \frac{e(\overline{\mu}_1-Q_0)}{2M^2} \bm{S} \cdot (\widehat{\bm{E}} \times \widehat{\bm{\pi}}) \pm
\frac{ie(\overline{\mu}_1-Q_0)}{4M^2} \bm{S} \cdot (\overline{\bm{\nabla}} \times \widehat{\bm{E}})
\nonumber\\
&& \mp \frac{\braket{r^2}_E}{6} \overline{\bm{\nabla}} \cdot \widehat{\bm{E}}~\mathbb{I}_{3\times3}
 \mp \frac{Q_2}{4}\left[S_iS_j+S_jS_i-\frac{2}{3} S^2\delta_{ij} \right]\overline{\bm{\nabla}}_i \widehat{\bm{E}}_j+
\mathcal{O}\left(\frac{1}{M^3},F^2\right).
\label{eq:H-pm-E}
\end{eqnarray}

The quantum-mechanical Green's functions of the theory, $\mathcal{G}^{(\pm)}_{\lambda,\lambda'}(\bm{x},t;\bm{x}',t')$, are defined to satisfy
\begin{eqnarray}
\left[ i\frac{d}{dt} \mp \widehat{\mathcal{H}}^{(\pm)}_{\text{NR}}(\widehat{\bm{\pi}},\widehat{\bm{x}}_3) \right]\mathcal{G}^{(\pm)}_{\lambda,\lambda'}(\bm{x},t;\bm{x}',t')
=i\delta^3(\bm{x}-\bm{x}')\delta(t-t')\delta_{\lambda,\lambda'},
\label{eq:GF-EOM-I}
\end{eqnarray}
for $\pm(t-t')>0$, and
\begin{eqnarray}
\mathcal{G}^{(\pm)}_{\lambda,\lambda'}(\bm{x},t;\bm{x}',t')=0,
\label{eq:GF-EOM-II}
\end{eqnarray}
for $\pm(t-t')<0$.\footnote{This choice of boundary conditions ultimately corresponds to the Feynman prescription for the propagator in the corresponding field theory description, as adopted in Sec. \ref{sec:Rel-Greens-functions}.} Subscripts $\lambda$ and $\lambda'$ refer to the polarization vectors (see below) associated with the wavefunction at points $(\bm{x},t)$ and $(\bm{x}',t')$, respectively. Since the electric charge radius and the electric quadrupole moment of the composite particle are already accessible through operators at $\mathcal{O}\left(\frac{1}{M^2}\right)$ in the NR Hamiltonian, it suffices to consider an electric field that has only a nonzero spatial gradient, as was the case in the previous section. With the choice of the gauge potential in Eq. (\ref{eq:A-field-NU}), the NR Hamiltonian does not depend on the $\bm{x}_2$ and $\bm{x}_3$ coordinates, and the conjugate momenta corresponding to these coordinates, namely $\bm{p}_1$ and $\bm{p}_2$, are conserved quantum numbers of the system. As in the relativistic case, the wavefunctions can be written in terms of plane waves along these directions. Consequently,
\begin{eqnarray}
\mathcal{G}^{(\pm)}_{\lambda,\lambda'}(\bm{x},t;\bm{x}',t')=\frac{1}{(2\pi)^2}
\int d\bm{p}_1 d\bm{p}_2~
e^{i\bm{p}_1(\bm{x}_1-\bm{x}_1')+i\bm{p}_2(\bm{x}_2-\bm{x}_2')}\widetilde{\mathcal{G}}^{(\pm)}_{\lambda,\lambda'}(\bm{x}_3,t;\bm{x}_3',t';\bm{p}_1,\bm{p}_2).
\label{eq:GF-plane-waves}
\end{eqnarray}
The dependence of $\widetilde{\mathcal{G}}^{(\pm)}$ on transverse momenta  $\bm{p}_1$ and $\bm{p}_2$ can in general be nontrivial because of the spin-orbit term in the NR Hamiltonian, but it is straightforward to explicitly work out the functional form of the solutions. In what follows, we choose to project onto the sector of zero transverse momentum, $\bm{p}_1=\bm{p}_2=0$, just as pursued in the previous section.\footnote{This projection removes the dependence of the NR Green's functions on the magnetic moment of the particle in an external electric field. However, the magnetic moment can be straightforwardly obtained form the Green's functions in a uniform magnetic field, and is not of primary interest in this paper.} The desired projection can be obtained by integrating both sides of Eq. (\ref{eq:GF-EOM-I}) over $\bm{x}_1$ and $\bm{x}_2$. The projected Green's functions then satisfy the following differential equation
\begin{eqnarray}
\left[ i\frac{d}{dt} \mp \widehat{\mathcal{H}}^{(\pm)}_{\text{NR}}(\widehat{\bm{x}}_3,\widehat{\bm{p}}_3) \right] \widetilde{\mathcal{G}}^{(\pm)}_{\lambda,\lambda'}(\bm{x}_3,t;\bm{x}_3',t';\bm{p}_1=\bm{p}_2=0)
=i\delta(\bm{x}_3-\bm{x}_3')\delta(t-t')\delta_{\lambda,\lambda'},
\label{eq:projected-GF-EOM-I}
\end{eqnarray}
for $\pm(t-t')>0$, and
\begin{eqnarray}
\widetilde{\mathcal{G}}^{(\pm)}_{\lambda,\lambda'}(\bm{x}_3,t;\bm{x}_3',t';\bm{p}_1=\bm{p}_2=0)
=0,
\label{eq:projected-GF-EOM-II}
\end{eqnarray}
for $\pm(t-t')<0$. In these equations,
\begin{eqnarray}
\widetilde{\mathcal{G}}^{(\pm)}_{\lambda,\lambda'}(\bm{x}_3,t;\bm{x}_3',t';\bm{p}_1=\bm{p}_2=0) \equiv
\int d\bm{x}_1 d\bm{x}_2 ~ \mathcal{G}^{(\pm)}_{\lambda,\lambda'}(\bm{x},t;\bm{x}',t'),
\label{eq:projected-GF}
\end{eqnarray}
and
\begin{eqnarray}
\widehat{\mathcal{H}}^{(\pm)}_{\text{NR}}(\widehat{\bm{x}}_3,\widehat{\bm{p}}_3)=\left(\frac{\widehat{\bm{p}}_3^2}{2M} \mp \frac{1}{2}M\omega_E^2\widehat{\bm{x}}_3^2 + M \mp \frac{E_0(\braket{r^2}_E+Q_2)}{6} \right) \mathbb{I}_{3 \times 3} \pm \frac{E_0Q_2}{2} \mathbb{J}_{3 \times 3},
\label{eq:H-NR-projected}
\end{eqnarray}
where $\omega_E^2 \equiv \frac{eQ_0E_0}{M}$ as before, and $\mathbb{J}$ is a $3 \times 3$ matrix whose only nonzero component is $(\mathbb{J})_{33}=1$. 

One way to to evaluate these Green's functions is to first solve for the wavefunctions in the background of this gauge potential. The wavefunctions in the zero transverse momentum sector, $\widetilde{\bm{\psi}}^{(\pm)}_{\text{NR}}$, satisfy the following Schr\"odinger equations
\begin{eqnarray}
i\frac{d}{dt}\widetilde{\bm{\psi}}^{(\pm)}_{\text{NR}}(\bm{x}_3,t)
= \pm \widehat{\mathcal{H}}^{(\pm)}_{\text{NR}}(\widehat{\bm{x}}_3,\widehat{\bm{p}}_3)\widetilde{\bm{\psi}}^{(\pm)}_{\text{NR}}(\bm{x}_3,t),
\label{eq:NRpsi-EOM-omega}
\end{eqnarray}
with $\widehat{\mathcal{H}}^{(\pm)}_{\text{NR}}(\widehat{\bm{x}}_3,\widehat{\bm{p}}_3)$ defined above. Depending on the sign of $eQ_0E_0$, Eq. (\ref{eq:NRpsi-EOM-omega}) is a Schr\"odinger  equation in the presence of either a normal or an inverted harmonic oscillator potential. As is well-known, the discrete eigenfunctions of the normal harmonic oscillator are a set of \emph{particular} solutions of the corresponding differential equation with boundary conditions $\widetilde{\psi}_{\text{NR}}(\bm{x}_3,t)=0~\text{as}~\bm{x}_3 \rightarrow \infty$ and $\int d\bm{x}_3 |\widetilde{\psi}_{\text{NR}}|^2 =\mathcal{N}$ for some finite constant $\mathcal{N}$. These conditions obviously do not hold for the wavefunction in an inverted oscillator potential. Nonetheless, at the level of the solutions to the corresponding Cauchy boundary-value problem, i.e., the \emph{fundamental} solutions, these are closely related by an analytic continuation in the oscillator frequency, $\omega_E \rightarrow i\omega_E$.\footnote{It would be wrong to perform this analytic continuation at the level of particular solutions. Not surprisingly, if this is done, the corresponding energy eigenvalues would be purely complex, $\mathcal{E}_n=i\omega_E(n+\frac{1}{2})$ for $n \in \mathbb{Z}$, which contradicts the fact that the Hamiltonian of the inverted harmonic oscillator is Hermitian.} Here we aim to find such general solutions. By performing a change of variables to $\xi \equiv \frac{ \bm{x}_3}{\cosh ( \omega_E t )}$ and $\vartheta \equiv \frac{\tanh ( \omega_E t )}{\omega_E}$ 
in Eq. (\ref{eq:NRpsi-EOM-omega}), the upper-component wavefunction $\widetilde{\bm{\psi}}_i^{(+)}$, with $i=1,2,3$ denoting the Cartesian-coordinate indices, can be written as
\begin{eqnarray}
\label{eq:psi-p-change-of-var-I}
\left(\widetilde{\bm{\psi}}^{(+)}_{\text{NR}}\right)_{1,2}&=&\frac{1}{\sqrt{\cosh ( \omega_E t )}} e^{-iMt+\frac{i}{2}  M\omega_E \tanh (\omega_E t ) \bm{x}_3^2+\frac{iE_0}{6}(\braket{r^2}_E+Q_2)t} \times u^{(+)}(\xi,\vartheta),
\\
\left(\widetilde{\bm{\psi}}^{(+)}_{\text{NR}}\right)_{3}&=&\frac{1}{\sqrt{\cosh ( \omega_E t )}} e^{-iMt+\frac{i}{2}  M\omega_E \tanh (\omega_E t ) \bm{x}_3^2+\frac{iE_0}{6}(\braket{r^2}_E-2Q_2)t} \times u^{(+)}(\xi,\vartheta),
\label{eq:psi-p-change-of-var-II}
\end{eqnarray}
where the function $u^{(+)}$ satisfies a free Schr\"odinger equation with respect to the new variables \cite{Polyanin:2002},
\begin{eqnarray}
i\frac{du^{(+)}}{d\vartheta}=-\frac{1}{2M}\frac{d^2u^{(+)}}{d \xi^2},
\label{eq:u-p-EOM}
\end{eqnarray}
and has the following plane-wave solution
\begin{eqnarray}
u^{(+)}(\xi,\vartheta)=e^{-i\frac{k^2}{2M}\vartheta+ik\xi},
\label{eq:u-p-sol}
\end{eqnarray}
with an arbitrary $k$. The subscripts on the wavefunctions denote the corresponding Cartesian components. Similarly, the solution of the wave equation (\ref{eq:NRpsi-EOM-omega}) for the lower-component wavefunction is\footnote{These solutions are easily obtained by replacing $t \rightarrow -t$ and $\omega_E \rightarrow i\omega_E$ in the positive-energy solutions, see Eqs. (\ref{eq:H-NR-projected}) and (\ref{eq:NRpsi-EOM-omega}).}
\begin{eqnarray}
\label{eq:psi-m-change-of-var-I}
\left(\widetilde{\bm{\psi}}^{(-)}_{\text{NR}}\right)_{1,2}&=&\frac{1}{\sqrt{\cos ( \omega_E t )}} e^{iMt+\frac{i}{2}  M\omega_E \tan (\omega_E t ) \bm{x}_3^2+\frac{iE_0}{6}(\braket{r^2}_E+Q_2)t} \times u^{(-)}(\eta,\theta),
\\
\left(\widetilde{\bm{\psi}}^{(-)}_{\text{NR}}\right)_{3}&=&\frac{1}{\sqrt{\cos ( \omega_E t )}} e^{iMt+\frac{i}{2}  M\omega_E \tan (\omega_E t ) \bm{x}_3^2+\frac{iE_0}{6}(\braket{r^2}_E-2Q_2)t} \times u^{(-)}(\eta,\theta),
\label{eq:psi-m-change-of-var-II}
\end{eqnarray}
where $\eta \equiv \frac{ \bm{x}_3}{\cos ( \omega_E t )}$ and $\theta \equiv \frac{\tan ( \omega_E t )}{\omega_E}$, 
and where the function $u^{(-)}$ satisfies  \cite{Polyanin:2002}
\begin{eqnarray}
i\frac{du^{(-)}}{d\theta}=\frac{1}{2M}\frac{d^2u^{(-)}}{d \eta^2},
\label{eq:u-m-EOM}
\end{eqnarray}
with the following plane-wave solutions
\begin{eqnarray}
u^{(-)}(\xi,t)=e^{i\frac{k^2}{2M}\theta+ik\eta},
\label{eq:u-m-sol}
\end{eqnarray}
where k is arbitrary. Out of these solutions, one can construct 6 independent, mutually orthogonal modes as following
\begin{eqnarray}
\widetilde{\bm{\psi}}^{(+)}_{(M_S)} \equiv 
\begin{pmatrix}
\mathcal{T}_{(M_S)}\widetilde{\bm{\psi}}^{(+)}_{\text{NR}} \\
   \mathbb{O}
 \end{pmatrix},
 ~~~\widetilde{\bm{\psi}}^{(-)}_{(M_S)} \equiv 
\begin{pmatrix}
\mathbb{O} \\
   \mathcal{T}_{(M_S)}\widetilde{\bm{\psi}}^{(-)}_{\text{NR}}
 \end{pmatrix},
 \label{eq:psi-all-modes}
\end{eqnarray}
where $\mathcal{T}_{(M_S)}$ matrices with $M_S=0,\pm 1$, project onto the three polarizations of a spherical tensor of rank $1$, 
\begin{eqnarray}
\mathcal{T}_{\left(M_S=-1\right)}= \frac{1}{\sqrt{2}} \begin{pmatrix}
    0 & 0 & 0 \\
    0 & 0 & 0 \\
    1 & -i & 0
 \end{pmatrix},
 ~\mathcal{T}_{\left(M_S=0\right)}= \begin{pmatrix}
    0 & 0 & 0 \\
    0 & 0 & 1 \\
    0 & 0 & 0
 \end{pmatrix},
 ~\mathcal{T}_{\left(M_S=1\right)}= \frac{1}{\sqrt{2}} \begin{pmatrix}
    -1 & -i & 0 \\
    0 & 0 & 0 \\
    0 & 0 & 0
 \end{pmatrix},
 \label{eq:psi-MS-polarizations}
\end{eqnarray}
and $\mathbb{O}$ is a null three-vector.
 
Having found these general solutions, it is now straightforward to construct the Green's functions.  For the modes consisting of nonzero upper components, one obtains
\begin{eqnarray}
\widetilde{\mathcal{G}}^{(+)}_{M_S,M_S'}(\bm{x}_3,t;\bm{x}_3',t';\bm{p}_1=\bm{p}_2=0)&=&
\sqrt{\frac{M\omega_E}{2 \pi i \sinh ( \omega_E t-\omega_Et' )}}
e^{\frac{iM\omega_E}{2 \sinh (\omega_Et-\omega_Et')}\left[\cosh (\omega_Et-\omega_Et')(\bm{x}_3^2+\bm{x}^{'2}_3)-2\bm{x}_3\bm{x}_3'\right]}
\nonumber\\
&& \hspace{1 cm} \times ~ e^{-iM(t-t')+\frac{iE_0}{6}(\braket{r^2}_E+a^{(M_S)}Q_2)(t-t')} ~\theta(t-t')~\delta_{M_S,M_S'},
\label{eq:NR-GF-upper}
\end{eqnarray}
which corresponds to the propagation of positive-energy modes forward in time, while for the modes with nonzero lower components,
\begin{eqnarray}
\widetilde{\mathcal{G}}^{(-)}_{M_S,M_S'}(\bm{x}_3,t;\bm{x}_3',t';\bm{p}_1=\bm{p}_2=0)&=&
-\sqrt{\frac{-M\omega_E}{2 \pi i \sin ( \omega_E t-\omega_Et' )}}
e^{-\frac{iM\omega_E}{2 \sin (\omega_Et-\omega_Et')}\left[\cos (\omega_Et-\omega_Et')(\bm{x}_3^2+\bm{x}^{'2}_3)-2\bm{x}_3\bm{x}_3'\right]}
\nonumber\\
&& \hspace{1 cm} \times ~ e^{iM(t-t')+\frac{iE_0}{6}(\braket{r^2}_E+a^{(M_S)}Q_2)(t-t')} ~\theta(t'-t) ~ \delta_{M_S,M_S'},
\label{eq:NR-GF-lower}
\end{eqnarray}
which corresponds to the propagation of negative-energy modes backward in time.  Additionally, the Green's functions between modes with positive and negative energies are vanishing as desired due to the orthogonality of the associated basis vectors. Here we have defined $a^{(M_S=\pm 1)}=1$ and $a^{(M_S=0)}=-2$. As expected, the electric field that is considered does not mix polarization states.

Finally, to make a connection to LQCD calculations, these Green's functions must be transformed to Euclidean spacetime, $t \rightarrow -i\tau$, and be made consistent with PBCs, 
\begin{eqnarray}
\widetilde{\mathcal{G}}^{(\pm),FV}_{M_S,M_S'}(\bm{x}_3+L,\tau+T;\bm{x}_3',\tau';\bm{p}_1=\bm{p}_2=0) =
 \widetilde{\mathcal{G}}^{(\pm),FV}_{M_S,M_S'}(\bm{x}_3,\tau;\bm{x}_3',\tau';\bm{p}_1=\bm{p}_2=0).
\label{eq:NR-GF-FV-BCs}
\end{eqnarray}
As already discussed, for systems that possess localized wavefunctions, the Euclidean FV Green's function can be simply written in terms of the infinite-volume Green's function. Explicitly,
\begin{align}
\widetilde{\mathcal{G}}^{(\pm),FV}_{M_S,M_S'}(\bm{x}_3,\tau;\bm{x}_3',\tau';\bm{p}_1=\bm{p}_2=0) =
\sum_{n,\nu}
 \widetilde{\mathcal{G}}^{(\pm)}_{M_S,M_S'}(\bm{x}_3+nL,\tau+\nu T;\bm{x}_3',\tau';\bm{p}_1=\bm{p}_2=0),
\label{eq:NR-GF-upper-lower-FV}
\end{align}
where $\nu$ and $n$ denote two integers, and as before, $T$ and $L$ refer to the temporal and spatial extents of the volume, respectively. Note that only nonnegative (nonpositive) integers contribute to the sum over $\nu$ in $\widetilde{\mathcal{G}}^{(+),FV}$ ($\widetilde{\mathcal{G}}^{(-),FV}$). Therefore, the sum over temporal images of the Green's functions remains bounded as $T \rightarrow \infty$, as long as $\frac{E_0}{6}(\braket{r^2}_E+a^{(M_S)}Q_2)$ is small compared to $M$. On the other hand, the condition of the wavefunctions being localized is only met for modes in a normal harmonic oscillator potential. For example, when positive-energy modes are considered, the external electric potential acts as a normal harmonic oscillator if $eQ_0E_0<0$, for which the FV Green's function can be obtained using Eq. (\ref{eq:NR-GF-upper-lower-FV}). For $eQ_0E_0<0$, this does not work and one should directly solve the Schr\"odinger equation in a finite volume with PBCs and match the result to the LQCD correlation function obtained in the same volume. This latter procedure, however, will not be necessary as all the low-energy parameters can be constrained in the former scenario, for which analytic knowledge of the solutions in the infinite volume is sufficient to construct the FV Green's functions in sufficiently large volumes.

In the following, the connection between these Green's functions and LQCD correlations functions will be studied further, and possible strategies to match the hadronic theory to LQCD calculations are introduced. We consider two such strategies in Secs. \ref{subsec:NR-transformed-correlators} and \ref{subsec:Landau-projection}. The first one is based on matching at the level of Green's functions. In this case, LQCD correlation functions are transformed to a suitable form to be directly compared with NR Green's functions of the effective hadronic theory in appropriate regions of spacetime. The reverse procedure can be also realized where the NR Green's functions are inversely transformed to correspond to lattice correlation functions. The second strategy considers matching at the level of energy eigenvalues for systems that possess energy eigenstates. In this method, the long-time behavior of correlation functions allows an extraction of the ground-state energies of the different polarizations of the system that can be compared with the energies obtained from the NR limit of the hadronic theory. Here, we focus specifically on an electric field that varies linearly in one spatial coordinate; for more general background fields other strategies may also be applicable.

\subsection{Matching at the level of Green's functions
\label{subsec:NR-transformed-correlators} 
}
To obtain the connection to LQCD correlation function, it is necessary to understand the relation between (semi-)relativistic Green's functions of Sec. \ref{sec:Rel-Greens-functions} and the NR Green's functions of this section. Such relation can be obtained by noting that EOM of the (semi-)relativistic Green's function,
\begin{eqnarray}
\left[i\frac{d}{dt}-\widehat{\mathcal{H}}_{\text{SR}}\right] G(\bm{x},t;\bm{x}',t')=i\delta^3(\bm{x}-\bm{x}')\delta(t-t'),
\label{eq:EOM-G-nonzero-gen}
\end{eqnarray}
can be brought to the following form
\begin{eqnarray}
&& \mathcal{U}(\bm{x},t;\widehat{\bm{\pi}})\left[i\frac{d}{dt}-\widehat{\mathcal{H}}_{\text{SR}}\right]\mathcal{U}^{-1}(\bm{x},t;\widehat{\bm{\pi}})\mathcal{U}(\bm{x},t;\widehat{\bm{\pi}})G(\bm{x},t;\bm{x}',t')\mathcal{U}^{-1}(\bm{x}',t';\widehat{\bm{\pi}}')
\nonumber\\
&& \hspace{5.625 cm}=\mathcal{U}(\bm{x},t;\widehat{\bm{\pi}})i\delta^3(\bm{x}-\bm{x}')\delta(t-t')\mathcal{U}^{-1}(\bm{x}',t';\widehat{\bm{\pi}}'),
\label{eq:EOM-G-nonzero-gen-trans}
\end{eqnarray}
by acting by a FWC transformation, $\mathcal{U}(\bm{x},t;\widehat{\bm{\pi}})$, from the left and an inverse transformation, $\mathcal{U}^{-1}(\bm{x}',t';\widehat{\bm{\pi}}')$, from the right. The discussions in the following are general, however, to give explicit expressions for the case of a time-independent electric field, we specify the FWC transformation up to the least order at which the NR Hamiltonian is sensitive to the charge radius and quadrupole moment. The transformation is
\begin{eqnarray}
&& \mathcal{U}(\bm{x},t,\widehat{\bm{\pi}})=e^{-i\mathcal{S}^{(2)}(\bm{x},t,\widehat{\bm{\pi}})}e^{-i\mathcal{S}^{(1)}(\bm{x},t,\widehat{\bm{\pi}})},
\end{eqnarray}
where $\mathcal{S}^{(1)}$ and $\mathcal{S}^{(2)}$ have been used in the previous section to reduce the relativistic Hamiltonian in Eq. (\ref{eq:rel-Hamiltonian-E}) to the NR Hamiltonian in Eq. (\ref{eq:Hamiltonian-nonzero-E}) in an external electric field. The transformation performed using $\mathcal{S}^{(1)}$ eliminates the odd terms (those proportional to $\sigma_1$ and $\sigma_2$) at $\mathcal{O}(\frac{1}{M})$ in the Hamiltonian, while that performed using $\mathcal{S}^{(2)}$ eliminates the odd terms at $\mathcal{O}(\frac{1}{M^2})$. These are explicitly given by
\begin{eqnarray}
\label{eq:S-1-def}
\mathcal{S}^{(1)}&=&\frac{i}{4M^2}\left[\widehat{\bm{\pi}}^2-2(\bm{S} \cdot \widehat{\bm{\pi}})^2\right]\sigma_1,
\\
\mathcal{S}^{(2)}&=&-\frac{ieC^{(0)}}{4M^3} \left[\bm{E} \cdot \widehat{\bm{\pi}}+\widehat{\bm{\pi}} \cdot \bm{E}-S_i S_j \bm{E}_j \widehat{\bm{\pi}}_i-S_i S_j  \widehat{\bm{\pi}}_j\bm{E}_i\right]  \sigma_2+\frac{eC^{(2)}_1}{2M^3} \overline{\bm{\nabla}} \cdot \bm{E} ~ \sigma_2
\nonumber\\
&& -\frac{eC^{(2)}_3}{2M^3}\left[ \overline{\bm{\nabla}} \cdot \bm{E}-\frac{1}{2}(S_iS_j+S_jS_i)\overline{\bm{\nabla}}_i\bm{E}_j \right] \sigma_2
-\frac{eQ_0}{8M^3}\left[ \widehat{\bm{\pi}}^2-2(\bm{S} \cdot \widehat{\bm{\pi}})^2,\varphi \right]\sigma_2,
\label{eq:S-2-def}
\end{eqnarray}
with $\bm{S}$ being defined in Eq. (\ref{eq:spin-matrices}) and $\sigma_i$s are the Pauli matrices. Note that the FWC transformation explicitly depends on parameters of the effective theory. Now returning to Eq. (\ref{eq:EOM-G-nonzero-gen-trans}), and by realizing that
\begin{eqnarray}
&& \widehat{\mathcal{H}}_{\text{NR}}=\mathcal{U}(\bm{x},\tau;\widehat{\bm{\pi}})\widehat{\mathcal{H}}_{\text{SR}}\mathcal{U}^{-1}(\bm{x},\tau;\widehat{\bm{\pi}}),
\end{eqnarray}
one arrives at
\begin{eqnarray}
&& \left[i\frac{d}{dt}-\widehat{\mathcal{H}}_{\text{NR}}\right]\mathcal{U}(\bm{x},\tau;\widehat{\bm{\pi}})G(\bm{x},t;\bm{x}',t')\mathcal{U}^{-1}(\bm{x}',\tau';\widehat{\bm{\pi}}')=i\delta^3(\bm{x}-\bm{x}')\delta(t-t').
\end{eqnarray}
This can be realized by integrating both sides of Eq. (\ref{eq:EOM-G-nonzero-gen-trans}) over $t'$ and $\bm{x}'$. Comparing this with the EOM of the NR Green's function, Eq. (\ref{eq:GF-EOM-I}), suggests that
\begin{eqnarray}
&& \mathcal{G}^{(\pm)}_{M_S,M_S'}(\bm{x},t;\bm{x}',t')=\mathcal{P}^{(\pm)}\otimes \mathcal{T}_{(M_S)}\left[\mathcal{U}(\bm{x},\tau;\widehat{\bm{\pi}})G(\bm{x},t;\bm{x}',t')\mathcal{U}^{-1}(\bm{x}',\tau';\widehat{\bm{\pi}}') \right] \mathcal{P}^{(\pm)}\otimes \mathcal{T}_{(M_S)}^{T}.
\label{eq:SR-GR-to-NR-GR}
\end{eqnarray}
$\mathcal{P^{(\pm)}}=\frac{1 \pm \sigma_3}{2}$ is an operator that projects onto the upper/lower three components of the 6-component wavefunction, $\psi$. After such projection, $\mathcal{T}_{(M_S)}$, as defined in Eq. (\ref{eq:psi-MS-polarizations}), transforms cartesian components of the upper- or lower-component wavefunctions to the components of a spherical tensor of rank 1. $\mathcal{T}_{(M_S)}^{T}$ denotes the transpose of matrix $\mathcal{T}_{(M_S)}$. However, Eq. (\ref{eq:SR-GR-to-NR-GR}) as written is misleading since although in the relativistic theory the propagation of modes when $|\bm{x}-\bm{x}'|$ is comparable to $|t-t'|$ is legitimate, in the NR theory, a NR speed of propagation, $v \equiv |\bm{x}-\bm{x}'|/|t-t'|$, requires $|\bm{x}-\bm{x}'| \ll |t-t'|$. As a result the equivalence between the transformed relativistic Green's functions and NR Green's functions in Eq. (\ref{eq:SR-GR-to-NR-GR}) can only be established upon realizing a small velocity expansion of the right-hand side of this equation. To illuminate this latter point, the example of noninteracting spin-1 Green's functions is studied in more detail in Appendix \ref{App:Rel-NR-relation}.

To be comparable to the NR Green's functions of the hadronic theory, the LQCD correlation function, as defined in Eq. (\ref{eq:correlator-def}), should be also transformed in a manner similar to a FWC transformation. Assuming that the dominant contribution to the correlation function in external fields arises from the single-particle state of the hadronic theory, and given the choice of the EM gauge potential in Eq. (\ref{eq:A-field-NU}), the desired transformation must act as
\begin{eqnarray}
\widetilde{C}^{(\pm)}_{M_S,M_S'}(\bm{x}_3,\tau;\bm{x}'_3,\tau')=\mathcal{P^{(\pm)}}\otimes \mathcal{T}_{(M_S)}~ \mathcal{U}(\bm{x}_3;\bm{p}_1=\bm{p}_2=0,\widehat{\bm{p}}_3)
\left[\sum_{\bm{x}_1,\bm{x}_2}C(\bm{x},\tau;\bm{x}',\tau')\right]
\nonumber\\
\mathcal{U}^{-1}(\bm{x}'_3;\bm{p}'_1=\bm{p}'_2=0,\widehat{\bm{p}}'_3) ~ \mathcal{P^{(\pm)}}\otimes \mathcal{T}_{(M_S)}^{T},
\label{eq:correlator-NR-trans-II}
\end{eqnarray}
where the correlation functions are projected to the zero transverse momentum sector by simply summing over the transverse coordinates. This form, upon a continuum extrapolation, can be directly compared with the FV NR Green's functions in Eq. (\ref{eq:NR-GF-upper-lower-FV}) in the NR regime. While in the infinite volume such region corresponds to $|\bm{x}-\bm{x}'| \ll |t-t'|$, in a finite volume with PBCs identifying this region requires further investigation. As is demonstrated in Appendix \ref{App:Rel-NR-relation}, for the weak background fields considered here, the NR region corresponds approximately to $|\bm{x}-\bm{x}'| \rightarrow 0,L$ and $|\tau-\tau'| \rightarrow T/2$, where $L$ and $T$ correspond to spatial and temporal extent of the volume, respectively. Finally, in order to match the transformed correlation functions in Eq. (\ref{eq:correlator-NR-trans-II}) to the projected Green's functions in Eq. (\ref{eq:NR-GF-upper-lower-FV}), it must be mentioned that, as noted above, the FWC transformation performed by the operator $\mathcal{U}$ is itself dependent upon the mass as well as the low-energy coefficients of the hadronic theory, $C^{(0)}$, $C^{(2)}_1$ and $C^{(2)}_3$ (or in turn the EM structure couplings, $\mu_1$, $\braket{r^2}_E$ and $Q_2$). As a result, to constrain these couplings, a rather elaborate fitting is required: first for each source location, the transformed correlation function at each point $\bm{x}_3$ and $\tau$ must be evaluated as a function of the low-energy parameters as well as the mass of the state. These can then be matched to the NR Green's functions with dependencies on the same parameters, which enables simultaneous constraint of the mass and all the EM couplings. Although this is not a simple fitting procedure, it appears to be more straightforward than a direct matching to semi-relativistic Green's functions of the previous section. 

Alternatively one can reverse the procedure described above and obtain the inverse FWC transform of the NR Green's functions. Then, up to the order at which one desires to keep terms in an $\frac{1}{M}$ expansion, the NR Green's functions will be promoted to a nondiagonal ``quasi-relativistic'' Green's function, which can be approximated by the relativistic Green's functions up to this order. As long as a NR limit of the LQCD correlation functions, corresponding to region of small velocity $v$, is considered, each component of the correlation function can be matched onto the (inversely) transformed NR Green's functions. This matching can prove simpler in practice than that discussed above. This is because transformation of numerical data is not required with transform functions that themselves depend upon the unknown parameters of the effective  theory.

We emphasize once again that the scope of validity of all the matching procedures described here and in the following is affected by the assumption of the single-particle dominance over the excited-states contributions to the correlation functions. This can be ensured to be a valid approximation by considering correlation functions at large Euclidean times, when the contamination from the excited states of the theory has died off compared with the single-particle state of interest. One can test whether this is the case by the goodness of fit for the above procedures; i.e., if multiple states are still contributing, a low goodness of fit will result. By focusing on the large-time behavior of correlation functions, one can also ensure the NR region of the correlation functions is being studied as required by the matching procedures of this section. At these large times,\footnote{Here and elsewhere, the large Euclidean time must be defined as the region where the contributions from backward propagating modes have not become significant, corresponding to $\tau-\tau' \rightarrow \frac{T}{2}$. This also corresponds to the proper NR region as discussed in Appendix \ref{App:Rel-NR-relation}.} one may additionally extract the energy eigenvalues (if they exist) for each polarization state of the system in its ground state in the presence of the weak EM field, and match their values to energy eigenvalues obtained from the hadronic theory. This method will be discussed in more detail in the following.

\subsection{Matching at the level of energy eigenvalues
\label{subsec:Landau-projection} 
}
To be concise, we focus on the positive-energy solutions of the NR theory in the chosen external field of Eq. (\ref{eq:E-field-NU}). In order for the positive-energy solutions of the EOM to possess quantized energy eigenvalues in the infinite volume, the system must be confined to a normal harmonic oscillator potential, for which the corresponding energy eigenstates will be the well-known Landau wavefunctions (these are analogous to the wavefunctions of a charged particle in a uniform magnetic field). As mentioned earlier, this scenario can only occur if $eQ_0E_0<0$, see Eq. (\ref{eq:H-NR-projected}). Being able to define energy eigenstates, one can isolate the contribution from each of these states to the Green's functions. As discussed above, the case $eQ_0E_0>0$ requires solving the hadronic system directly in a finite volume (in which case well-defined energy eigenvalues will exist), however, we do not consider this latter case further.

Let us assume that the volume is large compared with the intrinsic size of the harmonic oscillator, $L\sqrt{M|\omega_E|} \gg 1$, such that the sum over images in the FV Green's function in Eq. (\ref{eq:NR-GF-upper-lower-FV}) is dominated by the infinite-volume Green's function (that obtained from Eq. (\ref{eq:NR-GF-upper})). In this limit, the desired projection can be achieved by considering the infinite-volume Green's function. First note that according to \emph{Mehler's formula},
\begin{eqnarray}
\sum_{n=0}^{\infty}\frac{\rho^n}{2^nn!} \mathcal{H}_n(\xi) \mathcal{H}_n(\eta) e^{-\frac{1}{2}(\xi^2+\eta^2)}
=\frac{1}{\sqrt{1-\rho^2}}e^{-\frac{(1+\rho^2)(\xi^2+\eta^2)/2-2\xi\eta\rho}{1-\rho^2}},
\label{eq:Mehler}
\end{eqnarray}
the projected Green's function in Eq. (\ref{eq:NR-GF-upper}) for $\omega_E^2<0$ can be rewritten as\footnote{Set $\xi=\sqrt{M|\omega_E|}\bm{x}_3$, $\eta=\sqrt{M|\omega_E|}\bm{x}_3'$ and $\rho=e^{-i|\omega_E|(t-t')}$ in Mehler's formula.}
\begin{eqnarray}
&& \widetilde{\mathcal{G}}^{(+)}_{M_S,M_S'}(\bm{x}_3,t;\bm{x}_3',t';\bm{p}_1=\bm{p}_2=0) =
\theta(t-t')\delta_{M_S,M_S'}~\sqrt{\frac{M|\omega_E|}{\pi}}~e^{\frac{iE_0}{6}(\braket{r^2}_E+a^{(M_S)}Q_2)(t-t')}
\nonumber\\
&& \hspace{2.5 cm} \times ~ e^{-\frac{M|\omega_E|}{2}(\bm{x}_3^2+{\bm{x}'_3}^2)} \sum_{n=0}^{\infty}\frac{e^{-i(n+\frac{1}{2})|\omega_E| (t-t')}}{2^nn!}\mathcal{H}_n(\sqrt{M|\omega_E|}\bm{x}_3) \mathcal{H}_n(\sqrt{M|\omega_E|}\bm{x}_3'),
\label{eq:GF-sum-over-n}
\end{eqnarray}
where $\mathcal{H}_n(x)$ is the Hermite polynomial of order $n$. In this form, the Green's function can be readily projected onto the $n^{\text{th}}$ Landau level using the orthogonality relations of the Hermite polynomials. Explicitly, this projection is done by
\begin{eqnarray}
&& \widetilde{\mathcal{G}}^{(+),n}_{M_S,M_S'}(\bm{x}_3,t;\bm{x}_3',t';\bm{p}_1=\bm{p}_2=0) \equiv  \int d\bm{x}_3~ e^{-\frac{M|\omega_E|}{2}\bm{x}_3^2} \mathcal{H}_n(\sqrt{M|\omega_E}|\bm{x}_3) \widetilde{\mathcal{G}}^{(+)}_{M_S,M_S'}(\bm{x}_3,t;\bm{x}_3',t';\bm{p}_1=\bm{p}_2=0) 
\nonumber\\
&& \hspace{1.75 cm} = \theta(t-t') \delta_{M_S,M_S'}\sqrt{M|\omega_E|} e^{-\frac{M|\omega_E|}{2}{\bm{x}'_3}^2}  \mathcal{H}_n(\bm{x}_3') 
\times ~ e^{-i\left[(n+\frac{1}{2})|\omega_E|-\frac{E_0}{6}\left(\braket{r^2}_E+a^{(M_S)}Q_2\right)\right](t-t')}.
\label{eq:GF-n-projected}
\end{eqnarray}
In a finite volume, the integration region is limited to $-\frac{L}{2} \leq \bm{x}_3 \leq \frac{L}{2}$, however in the large-volume limit considered above, corrections to Eq. (\ref{eq:GF-n-projected}) will be exponentially suppressed in $L\sqrt{M|\omega_E|}$. From this equation, the NR energy of the particle in the $n^{\text{th}}$ Landau level is
\begin{eqnarray}
\mathcal{E}_n^{(M_S)}(E_0)
=(n+\frac{1}{2})|\omega_E|-\frac{E_0}{6}\left(\braket{r^2}_E+a^{(M_S)}Q_2\right),
\label{eq:NR-energy-n-Landau}
\end{eqnarray}
at leading order in the electric field strength. We recall that $a^{(M_S=\pm 1)}=1$ and $a^{(M_S=0)}=-2$.

Despite the NR case, the energy eigenfunctions of the relativistic theory in this external field do not possess a simple analytic form, and as was concluded, are closely related to the anharmonic oscillator wavefunctions. As a result, a direct projection of the relativistic Green's functions to the lowest lying energy eigenstates for $e\hat{Q}E_0<0$, similar to what was done above for the NR Green's functions, might prove challenging. Nonetheless, in order to eliminate the $\bm{x}_3$ dependence of the correlation functions, one may project the Green's functions by a suitable function of $\bm{x}_3$, which does not have to necessarily correspond to an energy eigenfunction (for example plane waves could be used). Assuming the system possesses discrete energy eigenstates in the chosen external field, it eventually asymptotes to its ground state, giving rise to a simple exponential fall off in the projected correlation functions at large Euclidean times. After subtracting the mass term, the extracted energy can be matched to the expectation for the lowest NR energy of the system, i.e., Eq. (\ref{eq:NR-energy-n-Landau}) with $n=0$. Unfortunately, this only leads to a constraint on a combination of the mean-squared charge radius and the quadrupole moment.

In order to constrain the quadrupole moment and the charge radius independently, one can form correlation functions that have a clear connection to the NR polarization vectors labeled by the $M_S$ quantum number. This can be achieved by performing the NR transformation of Eq. (\ref{eq:correlator-NR-trans-II}). This transformation not only decouples the upper and lower three components of the relativistic states in the two-point function, but also converts them to the convenient $M_S$ basis. The only complication is that such transformation contains the unknown low-energy parameters of the hadronic theory that are aimed to be extracted, see Eqs. (\ref{eq:S-1-def}) and (\ref{eq:S-2-def}). This is not ultimately a problem given that even an approximate transformation, such as the one performed with only the leading operator $\mathcal{S}^{(1)}$ in Eq. (\ref{eq:S-1-def}), already transforms the correlation function to that corresponding to the NR modes in the large Euclidean times (recalling that the background field is taken to be weak). Then, by separately forming the transformed correlation functions with longitudinal ($M_S=0$) and transverse ($M_S=\pm 1$) modes and subsequently performing a Landau-level projection on the transformed correlation function as in Eq. (\ref{eq:GF-n-projected}), the desired parameters of the hadronic theory can be constrained. Explicitly, one matches the energies obtained from the long-time behavior of the transformed correlation functions for each polarization to those of the hadronic theory, through which the mean-squared charge radius and the electric quadrupole moment will be constrained separately. For example, one may note that the spin-averaged energies, 
\begin{eqnarray}
\frac{1}{3}(\mathcal{E}_n^{(M_S=-1)}+\mathcal{E}_n^{(M_S=0)}+\mathcal{E}_n^{(M_S=1)})=(n+\frac{1}{2})|\omega_E|-\frac{E_0\braket{r^2}_E}{6},
\label{eq:E-isolated-r2}
\end{eqnarray}
does not depend on the quadrupole moment of the particle and will isolate the contribution from the charge radius. On the other hand, the difference in energies of the transverse and the longitudinal modes,
\begin{eqnarray}
\mathcal{E}_n^{(M_S=1)}+\mathcal{E}_n^{(M_S=-1)}-2\mathcal{E}_n^{(M_S=0)}=-E_0Q_2,
\label{eq:E-isolated-Q}
\end{eqnarray}
is insensitive to the charge radius and isolates the contribution due to the quadrupole moment. These results are reduced to those of particles with spin zero once $Q_2$ is set to zero, and can be used to extract the electric charge radius of scalar hadrons and nuclei. 

\subsection{On the extraction of deuteron's electric quadrupole moment and charge radius 
\label{subsec:Deuteron} 
}
Given our interest in constraining the electric quadrupole moment and the charge radius of the deuteron from QCD, and having obtained the shift in the NR energy of a composite spin-1 field in a linearly rising electric field in space, it is natural to ask how viable the extraction of such quantities is in upcoming LQCD calculations of this nucleus in the proposed background field. This is an important question in the light of the fact that the deuteron is a shallow bound state of two nucleons at physical value of quark masses, with a binding energy of $B=2.224644(34)~\text{MeV}$. Consequently, the range of validity of the single-particle description as adopted in this paper must be examined carefully to ensure that the applied external field does not resolve the internal structure of the bound state or does not break it up.

The condition of applicability of the single-particle formalism can be examined readily by comparing the zero-field binding energy of the deuteron, $B$, with the difference in the energy of the deuteron and that of the unbound neutron and proton in the chosen electric field, $\Delta \mathcal{E}^{(d,np)}$.\footnote{Here and in what follows, all energies must be realized as NR energies.} Since the experimental values of the electric charge radii (and the quadrupole moment for the case of the deuteron) are known for the proton, the neutron and the deuteron,\footnote{Discrepancies in the measured values of proton't charge radius will not matter here as we are only interested in a rough estimation of energy shifts..} one can estimate this energy difference for various electric-field slopes, $E_0$. First we note that the leading shift in the NR energy of a neutron, $\mathcal{E}^{(n)}(E_0)$ (with zero transverse momenta) due to interacting with the external $\mathbf{E}$ field of Eq. (\ref{eq:E-field-NU}) is
\begin{eqnarray}
\mathcal{E}^{(n)}(E_0)
=-\frac{E_0}{6}\braket{r_n^2}_E,
\label{eq:NR-energy-proton}
\end{eqnarray}
where $\braket{r_n^2}_E$ denotes the mean-squared electric charge radius of the neutron. The same energy shift for the case of proton, $\mathcal{E}_n^{(p)}(E_0)$, in its ground state is
\begin{eqnarray}
\mathcal{E}_n^{(p)}(E_0)
=\frac{|\omega_E^{p}|}{2}-\frac{E_0}{8M_p^2}-\frac{E_0}{6}\braket{r_p^2}_E,
\label{eq:NR-energy-proton}
\end{eqnarray}
where $\omega_E^{p}=\sqrt{\frac{eE_0}{M_p}}$, and $\braket{r_n^2}_E$ is the mean-squared electric charge radius of the proton.

To determine the slope of the linearly varying electric field for a viable extraction of the charge radius and the quadrupole moment, we assume that the upcoming LQCD calculations at the physical values of light-quark masses can be performed at volumes that are large compared with the intrinsic size of the deuteron, i.e., its inverse binding momentum, $\kappa$, resulting in small exponential corrections of the form $e^{-\kappa L}/L$ to the binding energy, where $L$ is the spatial extent of the volume. To be specific, let us take $L=17~\text{fm}$ such that the extracted ground-state energy in the absence of the background fields is within $10 \%$ of the infinite-volume deuteron binding energy \cite{Briceno:2013bda}. With a periodic implementation of the background field in Eq. (\ref{eq:E-field-NU}), the electric-field slope must be quantized as $eE_0=12 \pi n/L^2T$ with $n \in \mathbb{Z}$ \cite{Davoudi:2015cba}. As before, $T$ is the temporal extent of the volume which we take to be the same as $L$. In this scenario, the first quantum of the field slope results in an electric field that varies within $0 \leq |e\bm{E}| \leq 0.005~\text{GeV}^2$ throughout the volume. With this electric field, the difference between the (NR) ground-state energy of the deuteron in either of its polarization states and that of unbound proton and neutron, $\Delta \mathcal{E}^{(d,np)}$, will be negative (meaning that the threshold for the deuteron breakup is moved further away in this external field), rendering the effective single-particle description of this paper completely valid. This background field, however, results in a small contribution to the energy shift due to the particle's electric charge radius, see Eq. (\ref{eq:E-isolated-r2}), amounting for an energy shift that is only a few percent of the zero-field deuteron binding energy in this volume, making it challenging to isolate this contribution from the smallest quantum of the field slope (forming ratios of correlation functions may prove useful in such situations). On the other hand, the quadrupole moment contribution to the energy shift can be cleanly isolated from Eq. (\ref{eq:E-isolated-Q}), and results in a value that is $\sim 20 \%$ of the zero-field ground-state energy in this volume and may therefore be easier to isolate. The second and the third quanta of the electric-field slope make the values of quantities defined in Eqs. (\ref{eq:E-isolated-Q}) and (\ref{eq:E-isolated-r2}) more significant without pushing the limit of applicability of the single-particle description. 

Although the volume considered above allows us to input the physical values of the quadrupole moment and the charge radius for the deuteron for the purpose of estimating expected energy shifts, it is not a realistic scenario in practice in the near future.\footnote{Exponential corrections of the form $\sim  e^{-m_{\pi L}}/L$ to the mass as well as other low-energy parameters of the theory are considerably smaller than those corresponding to the size of the bound state and have therefore been neglected in the large volumes considered.} Such large volumes, however, may not be a necessity in precision studies of the deuteron. As long as the functional dependencies of quantities on the volume are known, their physical values can be extracted with high precision once fitted to these forms; a proposal that has been fully investigated for the case of the binding energy and scattering parameters in the deuteron channel in Ref. \cite{Briceno:2013bda}. In the same manner, a low-energy effective theory of the deuteron, for example the pionless EFT, can be used to determine how the charge radius and the quadruple moment behave as a function of volume. These as well as the knowledge of the volume dependence of the binding energy, can be used to extract the physical values of these quantities from LQCD calculations performed at significantly smaller volumes. As an example, a calculation performed at $L=9~\text{fm}$, with an associated zero-field ground-state energy of $\mathcal{E}=-B^{V} \approx -4.75 ~ \text{MeV}$, can result in significantly larger energy shifts in Eqs. (\ref{eq:E-isolated-r2}) and (\ref{eq:E-isolated-Q}). At this volume, however, even the smallest field quantum results in a field that pushes the FV ground-state energy toward the threshold, and will therefore violates the weak-field assumption. While still retaining the periodicity of the FV calculation, the need for quantized field slopes can be avoided  by shifting the point at which the electric field vanishes from $\bm{x}_3=0$ to $\bm{x}_3=L/2$, as shown in Ref. \cite{Davoudi:2015cba}. Then, one can take, for example, a half integer value for the field slope quanta, $E_0$, to allow the field to lie, at most, within $0 \leq |e\bm{E}| \leq 0.009~\text{GeV}^2$. This leaves the ground-state energy of the system at least $\sim 4~\text{MeV}$ away from threshold in this background field. With this choice, the value of the quantity defined in Eq. (\ref{eq:E-isolated-Q}) that is formed to isolate the quadrupole contribution will be $\sim 30 \%$ of the zero-field ground-state energy in this volume, and is likely to be extracted if a precision determination of energies is achieved. Note that these estimates are based on the infinite-volume values of the quadruple moment and the charge radius and must be considered to be only approximate. For the charge radius contribution, see Eq. (\ref{eq:E-isolated-r2}), the corresponding shift remains small, and is only $\sim 5 \%$ of the zero-field ground-state energy of the system. It is likely that by forming the energy differences and/or by constructing ratios of correlation functions at different values of background field slopes, significant improvements could be achieved in associated uncertainties (as observed in previous calculations in uniform background fields), allowing the extraction of both the quadruple moment and the charge radius.   

Before concluding this section, it is worth mentioning that at larger values of light-quark masses, the deuteron is rather deeply bound, as concluded by several LQCD collaborations, see e.g., \cite{Yamazaki:2015nka} for a review. This means that the validity of the single-particle theory is  guaranteed for a wider range of background-field strengths. Additionally, the volume corrections  due to the size of the bound state are largely suppressed and the extracted quantities will be close to their physical values at moderate volumes. The estimations made above for the energy shifts that are sensitive to the charge radius and the quadrupole moment are however not possible as the values of these quantities are unknown. Indeed, it will be an interesting task for LQCD to constrain these quantities for the deuteron at heavier quark masses. This will further complete the description of the properties of this nucleus in a world that differs from the physical world in the value of the input parameters of QCD.

\section{Conclusion
\label{sec:Conclusion} 
}
\noindent
One way to access electromagnetic (EM) properties of hadronic systems from first principles is to introduce classical EM fields into lattice quantum chromodynamics (LQCD) calculations. In the weak-field limit, the low-energy properties of hadrons and nuclei, such as their EM moments and polarizabilities, can be obtained from the response of the system to these external fields, and are characterized by an effective low-energy Hamiltonian. To extend the previous determinations of magnetic moments and polarizabilities of hadrons \cite{Bernard:1982yu, Martinelli:1982cb, Fiebig:1988en, Christensen:2004ca, Lee:2005ds, Lee:2005dq, Detmold:2006vu, Aubin:2008qp, Detmold:2009dx, Detmold:2010ts, Primer:2013pva, Lujan:2014kia} and light nuclei \cite{Beane:2014ora, Chang:2015qxa} to the case of charge radii and quadrupole moment, \emph{nonuniform} background fields must be implemented in LQCD calculations, as discussed in Ref. \cite{Davoudi:2015cba}. Additionally, the relationship between the EM couplings and LQCD correlation functions must be deduced. With special interest in the case of the deuteron, as well as (stable) vector mesons, we have obtained such relationships in this work for generic composite spin-1 fields in external EM fields. A general semi-relativistic effective field theory of spin-1 fields coupled to external electric and magnetic fields, including nonminimal couplings at $\mathcal{O}\left(\frac{Q^2}{M^2}\right)$, is presented, where $M$ is the mass of the spin-1 field and $Q$ is a typical momentum in the system and/or the inverse length scale over which the external field varies. The coefficients of the interactions are constrained by matching to on-shell amplitudes at low momentum transfer. The nonrelativistic (NR) reduction of the equations of motion leads to  effective Hamiltonians describing the response of the field to external EM fields in a systematic expansion in the inverse mass of the particle as well as its compositeness scale. The relations for the case of spin-0 hadrons and nuclei can be readily deduced from those obtained in this paper for the case of spin-1 fields. 

To provide explicit results that can be directly used in matching the effective hadronic theory to future LQCD calculations, we have chosen an electric field that varies linearly over space in one direction, and have obtained the corresponding Green's functions. Due to their involved form, it is potentially more practical to perform the matching at the level of the Green's functions of the NR theory. This requires applying a suitable NR transformation to lattice correlation functions. The NR Green's functions are explicitly obtained and the corresponding transformation of the correlation functions for the case of the electric field chosen in this work is presented. Alternatively, upon tuning the sign of the electric-field slope, the NR Green's functions can be projected onto Landau energy eigenstates. The energy eigenvalues corresponding to longitudinal and transverse modes of the NR theory are shown to have different dependencies on the quadrupole moment and the electric charge radius, providing the opportunity to constrain these quantities separately. By performing the same projection onto the Landau levels, the  transformed correlation functions at large Euclidean times exponentially asymptote to the projected Landau level, and the extracted energy can be matched to the NR energies obtained in the hadronic theory. Obtaining the exponential fall off of the transformed correlation functions in this way is potentially more practical. In particular, considering the fact that at large Euclidean times, the contamination from excited states of the theory is suppressed, and that the effective hadronic theory presented in this paper does not incorporate such excited states, the matching between the hadronic theory and LQCD is only reliable when such excitations of the hadron or nucleus are absent in the correlation functions.  Although in this paper we have focused on matching to LQCD correlation functions, with the background-field methodology, the ideas presented here may be applicable to nuclear many-body calculations, enabling an extraction of charge radii and electric quadrupole moments of nuclei.

Finally, the mass, the quadrupole moment and the charge radius that are constrained by the matching procedure must be extrapolated to infinite volume to correspond to their physical values. In the absence of a numerical extrapolation to infinite volume, analytic relations must be sought to determine the volume dependence of the low-energy couplings of the single-particle theory in a finite volume. These relations can only be obtained if one goes beyond the single-particle description presented in this work, and examines the ``around-the-world effects'' of the internal modes through a suitable low-energy effective theory. In the case of multi-hadron bound states such as the deuteron, this requires accounting for the finite-volume effects in a multi-hadron theory (e.g., the pionless effective field theory (EFT) in the case of the deuteron). Such studies must be conducted on a case-by-case basis as has been already done for several single-hadron properties using chiral perturbation theory \cite{Detmold:2006vu, Tiburzi:2008pa}, but must be extended to include properties such as charge radii and higher EM moments specially in multi-hadron cases.\footnote{See Refs. \cite{Bunton:2006va, Tiburzi:2014yra} for a discussion of the volume extrapolation of the  pseudo-scalar meson form factors within chiral perturbation theory.} For the case of the deuteron, one may alternatively formulate the problem in terms of the effective degrees of the freedom in the system, the nucleons, within an appropriate EFT that not only accounts for the interactions of the fields with the external EM field but also the internal interactions among the nucleons. Such approach has been taken for example in Ref. \cite{Detmold:2004qn} to constrain the low-energy constants of an EFT of the two-nucleon systems at low energies (the pionless EFT) in the presence of uniform background EM and weak fields. Such a formulation will also systematically account for the leading volume corrections that are not accounted for in the single-particle approach. Nonetheless, the single-particle description presented in this work is valid for the case of the deuteron for careful choices of the background field parameters, and as shown in Sec. \ref{subsec:Deuteron}, will make the extraction of the deuteron's charge radius and quadruple moment feasible, especially at larger values of quark masses where the volume corrections due to the size of the bound state are largely suppressed.

We have recently presented an implementation of general nonuniform EM fields on a periodic hypercubic lattice in Ref. \cite{Davoudi:2015cba}. By requiring the background gauge fields to respect the periodicity of the lattice calculation, one obtains a set of quantization conditions that must be satisfied by the back-ground field parameters. These nonuniform background fields, once implemented, can be used to obtain a broader range of EM properties than the ones considered in this paper. Some examples are the spin polarizabilities of hadrons \cite{PhysRevD.47.3757} for which some first attempts have been made previously \cite{Lee:2011gz, Engelhardt:2011qq}. In order to prevent possible inconsistencies when matching NR effective theories in background fields to on-shell processes (see Refs. \cite{Lee:2013lxa, Lee:2014iha}), one needs to start from the most general effective description that does not make unwarranted assumptions about the contributions from the equation-of-motion operators, as has been detailed in Refs. \cite{Lee:2013lxa, Lee:2014iha} for the case of spin-0 and spin-$\frac{1}{2}$ fields. These references, however, do not account for nonuniformities in the background fields. The formalism presented in this paper for spin-1 fields outlines a consistent approach when general background fields are considered. However, we have excluded contributions in the EM field-strength squared and higher, which is the order where  potential inconsistencies arise in isolating polarizabilities. An immediate extension of this work and those of Refs. \cite{Lee:2013lxa, Lee:2014iha} is to account for these terms in the background of nonuniform fields in the case of particles with arbitrary spin.

In summary, given the recent progress in LQCD calculations with the background field method, and with further formal developments similar to the study presented in this work, upcoming LQCD calculations will be able to constrain a wide range of EM properties of hadrons and nuclei, from their charge radii and higher EM moments to their EM form factors \cite{Detmold:2004kw}.

\
\

\begin{center}
\textbf{Acknowledgments}
\end{center}
\noindent
We would like to thank Michael G. Endres, Andrew Pochinsky and Jesse Thaler for interesting discussions, and Brian C. Tiburzi for fruitful correspondence. We are grateful to Martin J. Savage for discussions and for valuable comments on an earlier manuscript of this paper. ZD was supported by the U.S. Department of Energy under grant contract number DE-SC0011090. WD was supported by the U.S. Department of Energy Early Career Research Award DE-SC0010495 and by the National Science Foundation under Grant No. NSF PHY11- 25915. ZD and WD acknowledge the Kavli Institute for Theoretical Physics for hospitality during completion of this work.

\appendix
\section{On the Gauge Dependency of the Relativistic Green's Functions in External Fields
\label{App:Gauge-dependency}
}
\noindent
As was shown in Sec. \ref{sec:Rel-Greens-functions}, the evaluation of the (projected) semi-relativistic Green's function of composite spin-1 fields in a linearly varying electric field in space reduced to evaluating a simpler Green's function for structureless spin-0 fields,
\begin{eqnarray}
\widetilde{G}^{(0)}_{scl}(\bm{x}_3,t;\bm{x}'_3,t') = \int_0^{\infty} ds~ e^{-i(M^2-i\epsilon)s} \int_{-\infty}^{\infty} \frac{dp_0}{2\pi}~ e^{-ip_0(t-t')+ip_0^2s} \braket{\bm{x}_3|e^{-i\widehat{\mathcal{H}}^{(\bm{E})}s}|\bm{x}'_3}.
\label{eq:GF-0-rel-scalar-II-App}
\end{eqnarray}
For brevity, throughout this appendix the reference to the values of transverse momenta, $\bm{p}_1=\bm{p}_2=0$, is dropped in the argument of $\widetilde{G}^{(0)}_{scl}$. Clearly, this Green's function depends upon the choice of the gauge potential in Eq. (\ref{eq:A-field-NU}),
\begin{eqnarray}
A_{\mu}^{(\text{NU},1)}=(\varphi^{(\text{NU},1)},-\bm{A}^{(\text{NU},1)})=(-\frac{1}{2}E_0\bm{x}_3^2,\bm{0}),
\label{eq:A-field-NU-App}
\end{eqnarray}
and would have a different form once other choices of the gauge potential are considered. Here, the superscript $(\text{NU},1)$ is introduced to distinguish this gauge from a second choice below, and NU refers to the fact that a nonuniform electric field results from this potential. The aim of this appendix is to show that the Green's function in Eq. (\ref{eq:GF-0-rel-scalar-II-App}) can be separated into a gauge-dependent phase factor and a gauge-independent function whose corresponding EOM, despite the absence of an analytic solution for the Green's function, can be inferred straightforwardly. The separation of the gauge-dependent and gauge-independent parts of the Green's functions in external fields is well-known and arises naturally in the proper-time approach. In his pioneering paper, Schwinger considered the cases of a constant and a plane-wave EM fields, for which he was able to provide analytic solutions for the relativistic Green's functions of spin-$\frac{1}{2}$ fields \cite{Schwinger:1951nm}. These Green's functions exhibited the above feature, and the gauge-dependent phase factor in both cases was shown to be
\begin{eqnarray}
\Phi(x,x')=e^{-ie Q_0 \int_{\mathcal{C}(x,x')}d\xi_{\mu}A^{\mu}(\xi)}.
\label{eq:Phi-def-path}
\end{eqnarray}
$\mathcal{C}(x,x')$ in the exponent denotes the path over which the line integral is performed, and should be taken to be the straight line that starts from point $x'$ and ends at point $x$.\footnote{Any  path that produces the same result as that of the straight line is also acceptable.} Since we are limited by the lack of analytic results in Sec. \ref{sec:Rel-Greens-functions} (owing to a quartic potential), we deduce the separated form of the Green's function in Eq. (\ref{eq:GF-0-rel-scalar-II-App}), including the gauge dependent-phase factor, in an indirect manner.

To demonstrate the idea, it is useful to first discuss the case a spin-0 particle in a constant magnetic field for which an analytic solution exists. The analogies between this case and the nonuniform field considered in Sec. \ref{sec:Rel-Greens-functions} makes the deduction of the corresponding results in the latter case straightforward. To generate a constant magnetic field, $B$, along the $\bm{x}_3$ direction, one can choose an EM gauge potential of the form
\begin{eqnarray}
A^{(\text{U},1)}_{\mu}=(\varphi^{(\text{U},1)},-\bm{A}^{(\text{U},1)})=\left(0,B\bm{x}_2,0,0\right).
\label{eq:A-U-choice-I}
\end{eqnarray}
Superscript $(\text{U},1)$ has been introduced to distinguish this gauge from a second choice below, and $\text{U}$ refers to the fact that a uniform magnetic field is resulted from this potential. The relativistic Green's function of a structureless spin-0 particle in this background gauge potential satisfies
\begin{eqnarray}
\left[(D^{(\text{U},1)}_{\mu})^2+M^2\right]G^{(\text{U},1)}(x,x')=-i\delta^4(x-x').
\end{eqnarray}
$D^{(\text{U},1)}_{\mu}$ denotes the covariant derivative in this gauge, $D^{(\text{U},1)}_{\mu}=\partial_{\mu}+ieQ_0A^{(\text{U},1)}_{\mu}$, and all derivatives here, and in what follows, are taken with respect to the unprimed coordinates, $x$. Following the Schwinger method, it is straightforward to show that 
\begin{eqnarray}
&& G^{(\text{U},1)}(x,x')=\frac{1}{2} \int_{0}^{\infty} \frac{ds}{2 \pi s}   e^{-\frac{iM^2s}{2}-\frac{(\bm{x}_3-\bm{x}_3')^2-(t-t')^2}{2is}} \int \frac{d\bm{k}_1}{2\pi} e^{i\bm{k}_1(\bm{x}_1-\bm{x}_1')} \times 
\nonumber\\
&& \qquad ~~~ \sqrt{\frac{e Q_0B}{2\pi i \sin(eQ_0Bs)}} e^{\frac{ieQ_0B}{2\sin(eQ_0Bs)}\left\{ \cos(eQ_0Bs) \left[(\bm{x}_2+\frac{\bm{k}_1}{eQ_0B})^2+(\bm{x}'_2+\frac{\bm{k}_1}{eQ_0B})^2 \right]-2(\bm{x}_2+\frac{\bm{k}_1}{eQ_0B})(\bm{x}'_2+\frac{\bm{k}_1}{eQ_0B}) \right\}},
\label{eq:GF-gauge-I}
\end{eqnarray}
which implies that the system is a quantum harmonic oscillator in the $\bm{x}_2$ direction, while in all other directions, it has plane-wave solutions. This statement is clearly gauge dependent. In order to make the gauge dependency of $G^{(\text{U},1)}(x,x')$ more transparent, one can perform the integral over $\bm{k}_1$, after which Eq. (\ref{eq:GF-gauge-I}) turns into
\begin{eqnarray}
G^{(\text{U},1)}(x,x') &=& \Phi^{(\text{U},1)}(x,x')\mathbb{G}^{(U)}(x,x')
\label{eq:Phi-G},
\end{eqnarray}
where
\begin{eqnarray}
\Phi^{(\text{U},1)}(x,x')=e^{-\frac{i}{2}eQ_0B (\bm{x}_1-\bm{x}_1')(\bm{x}_2+\bm{x}_2')},
\label{eq:Phi-U-1}
\end{eqnarray}
is nothing but the phase factor introduced in Eq. (\ref{eq:Phi-def-path}) when the choice of $A^{(\text{U},1)}_{\mu}$ in Eq. (\ref{eq:A-U-choice-I}) is used. Further,
\begin{eqnarray}
\mathbb{G}^{(U)}(x,x')=
 i\int_{0}^{\infty} \frac{ds}{(4 \pi i s)^2} \frac{e Q_0Bs}{\sin(eQ_0Bs)}  e^{-iM^2s-\frac{(\bm{x}_3-\bm{x}_3')^2-(t-t')^2}{4is}+\frac{ieQ_0B}{4\tan(eQ_0Bs)} \left[(\bm{x}_1-\bm{x}_1')^2+(\bm{x}_2-\bm{x}'_2)^2 \right]},
\label{eq:gauge-inv-G}
\end{eqnarray}
is a translationally-invariant contribution to the Green's function \cite{Tiburzi:2014zva, Kuznetsov:2015uca}. This result is the direct consequence of the following identity,
\begin{eqnarray}
\left[(D_{\mu}^{(\text{U},1)})^2+M^2\right]G^{(\text{U},1)}(x,x') 
= \Phi^{(\text{U},1)}(x,x') \left[(D_{\mu}^{(\text{U},2)})^2+M^2\right]\mathbb{G}^{(U)}(x,x'),
\label{eq:identity-U}
\end{eqnarray}
with
\begin{eqnarray}
\left[(D_{\mu}^{(\text{U},2)})^2+M^2\right]\mathbb{G}^{(U)}(x,x')=-i\delta^4(x-x').
\label{eq:EOM-U-2}
\end{eqnarray}
$D^{(\text{U},1)}$ is defined above and $D_{\mu}^{(\text{U},2)}$ can be interpreted as a covariant derivative, $D_{\mu}^{(\text{U},2)}=\partial_{\mu}+ieQ_0A_{\mu}^{(\text{U},2)}$, in the following gauge
\begin{eqnarray}
A^{(\text{U},2)}_{\mu}=(\varphi^{(\text{U},2)},-\bm{A}^{(\text{U},2)})=\left(0,\frac{1}{2}B(\bm{x}_2-\bm{x}_2'),-\frac{1}{2}B(\bm{x}_1-\bm{x}_1'),0\right).
\label{eq:A-U-choice-II}
\end{eqnarray}
Although the identity in Eq. (\ref{eq:identity-U}) is deduced by starting with the EOM of the Green's function with the gauge potential in Eq. (\ref{eq:A-U-choice-I}), it in fact holds for any gauge that transforms to $A^{(\text{U},1)}_{\mu}$ via a gauge transformation.\footnote{This follows from the fact that $D_{\mu}\Phi(x,x')$ is gauge invariant as was shown in the Schwinger's original paper \cite{Schwinger:1951nm}. This also explains why one only needs to study the gauge dependency of the spin-$0$ piece of the Green's function in Eq. (\ref{eq:GF-0-rel-III}) as the accompanying terms, being a function of the covariant derivate, do not introduce additional gauge dependencies in the Green's functions.} As a result, Eq. (\ref{eq:GF-gauge-I}) has been separated into a gauge-variant phase factor, $\Phi$, and a gauge-invariant function, $\mathbb{G}$. 
Note that using Eq. (\ref{eq:Phi-def-path}), the phase factor corresponding to the gauge choice in Eq. (\ref{eq:A-U-choice-II}) evaluates to one as expected,
\begin{eqnarray}
\Phi^{(\text{U},2)}(x,x')&=&e^{-\frac{ie Q_0B}{2} \int_{\mathcal{C}(x,x')} [ (\bm{\xi}_2-\bm{x}'_2)d\bm{\xi}_1-(\bm{\xi}_1-\bm{x}_1')d\bm{\xi}_2 ]}
\nonumber\\
&=&e^{-\frac{ie Q_0B}{4}\left[ (\bm{x}_1-\bm{x}_1')(\bm{x}_2-\bm{x}'_2)-(\bm{x}_1-\bm{x}_1')(\bm{x}_2-\bm{x}'_2) \right]}=1,
\end{eqnarray}
where $\mathcal{C}(x,x')$ denotes the straight-line path as before. Therefore, the problem of finding the separated Green's functions is reduced to that of finding the gauge in which the phase factor, as defined in Eq. (\ref{eq:Phi-def-path}), is equal to one.

Returning to the case of the nonuniform electric field of Sec. \ref{sec:Rel-Greens-functions}, it is expected that the phase factor that carries the gauge dependency of the Green's functions in Eq. (\ref{eq:GF-0-rel-scalar-II-App}) will arise upon integration over $p_0$, in analogy with the case of a constant magnetic field. Although we do not know the analytic form of the Green's function in this case to directly perform the integral, we can deduce the separated Green's function by transforming to a gauge for which $\Phi(x,x')$ is equal to unity. Note that a similar identity as in Eq. (\ref{eq:identity-U}) can be written for the nonuniform case considered here,
\begin{eqnarray}
&&\left[(D_0^{(\text{NU},1)})^2-(D_3^{(\text{NU},1)})^2+M^2\right]\widetilde{G}^{(0)}_{scl}(\bm{x}_3,t;\bm{x}'_3,t') 
\nonumber\\
&&\qquad \qquad \qquad = \Phi^{(\text{NU},1)}(\bm{x}_3,t,\bm{x}'_3,t') \left[(D_0^{(\text{NU},2)})^2-(D_3^{(\text{NU},2)})^2+M^2\right]\widetilde{\bm{\mathbb{G}}}^{(0)}_{scl}(\bm{x}_3,t;\bm{x}'_3,t'),
\label{eq:identity-NU}
\end{eqnarray}
where $D_{\mu}^{(\text{NU},1)}$ is the covariant derivative defined with the gauge potential in Eq. (\ref{eq:A-field-NU-App}), and $D^{(\text{NU},2)}_{\mu}$ can be interpreted as the covariant derivate in the following gauge, 
\begin{eqnarray}
A^{(\text{NU},2)}_{\mu}=(\varphi^{(\text{NU},2)},-\bm{A}^{(\text{NU},2)})=\left( -\frac{1}{6}E_0(\bm{x}_3-\bm{x}_3')(2\bm{x}_3+\bm{x}_3'),0,0,\frac{1}{6}E_0(t-t')(2\bm{x}_3+\bm{x}_3') \right).
\end{eqnarray}
Now given the EOM for $\widetilde{G}^{(0)}_{scl}(\bm{x}_3,t;\bm{x}'_3,t')$ and the form of $\Phi^{(\text{NU},1)}(\bm{x}_3,t,\bm{x}'_3,t')$ that can be obtained from Eq. (\ref{eq:Phi-def-path}) (see below), the EOM in this gauge can be deduced from Eq. (\ref{eq:identity-NU}),
\begin{eqnarray}
\left[(D_0^{(\text{NU},2)})^2-(D_3^{(\text{NU},2)})^2+M^2\right]\widetilde{\bm{\mathbb{G}}}^{(0)}_{scl}(\bm{x}_3,t;\bm{x}'_3,t')=-i\delta(t-t')\delta(\bm{x}_3-\bm{x}_3'),
\label{eq:EOM-NU}
\end{eqnarray}
implying the phase factor corresponding to this gauge is unity. Clearly, this could be already inferred by  directly evaluating the line integral in Eq. (\ref{eq:Phi-def-path}) with this gauge choice,
\begin{eqnarray}
\Phi^{(\text{NU},2)}(\bm{x}_3,t,\bm{x}'_3,t')&=&e^{-\frac{ie Q_0E_0}{6} \int_{\mathcal{C}(x,x')} [ -(\bm{\xi}_3-\bm{x}'_3)(2\bm{\xi}_3+\bm{x}'_3)d\xi_0+(\xi_0-t')(2\bm{\xi}_3+\bm{x}'_3)d\bm{\xi}_3 ]}
\nonumber\\
&=&e^{-\frac{ie Q_0E_0}{36}\left[ -(t-t')(\bm{x}_3-\bm{x}'_3)(4\bm{x}_3+5\bm{x}'_3)+(t-t')(\bm{x}_3-\bm{x}'_3)(4\bm{x}_3+5\bm{x}'_3) \right]}=1.
\end{eqnarray}
Solving for the Green's function in this form is even more cumbersome due to the additional time-dependence of the (proper-time) Hamiltonian. However, we have succeeded in separating the gauge-dependent and gauge-independent parts of the Green's function corresponding to the simpler gauge choice of Eq. (\ref{eq:A-field-NU-App}). Explicitly,
\begin{eqnarray}
\widetilde{G}^{(0)}_{scl}(\bm{x}_3,t;\bm{x}'_3,t') = \Phi^{(\text{NU},1)}(\bm{x}_3,t,\bm{x}'_3,t') \widetilde{\bm{\mathbb{G}}}^{(0)}_{scl}(\bm{x}_3,t;\bm{x}'_3,t'),
\end{eqnarray}
where $\Phi(\bm{x}_3,t,\bm{x}'_3,t')$ can be evaluated using Eqs. (\ref{eq:Phi-def-path}) and (\ref{eq:A-field-NU-App}),
\begin{eqnarray}
\Phi^{(\text{NU},1)}(\bm{x}_3,t,\bm{x}'_3,t')=e^{\frac{i}{6}eQ_0E_0(t-t')(\bm{x}_3^2+\bm{x}_3\bm{x}_3'+{\bm{x}_3'}^{2})}.
\end{eqnarray}
An important feature to notice is that a nonuniform EM field breaks the translational invariance. So although it is possible to isolate the gauge-dependent part of the Green's function in Eq. (\ref{eq:GF-0-rel-scalar-II-App}), the remaining part, i.e., the $ \mathbb{G}(x,x')$ function, is not invariant under a translation in the coordinate in which the electric field varies, i.e., the $\bm{x}_3$ coordinate for the case considered here; a feature that is indeed observed from the EOM of  $\widetilde{\bm{\mathbb{G}}}^{(0)}_{scl}$, Eq. (\ref{eq:EOM-NU}).

To conclude this section, let us emphasize that even though the Green's functions are gauge dependent, once matched to LQCD correlation functions evaluated in the same background gauge fields, the obtained physical quantities will not be gauge dependent. Therefore, for practical applications, one does not need to worry about isolating the gauge dependence of the Green's functions in the hadronic theory as was done in this appendix.

\section{The relation between relativistic and nonrelativistic Green's functions of the spin-1 effective theory in the absence of external EM fields
\label{App:Rel-NR-relation}
}
To manifest the connection between the relativistic Green's function of Sec. \ref{sec:Rel-Greens-functions} and the NR Green's functions of Sec. \ref{sec:NR-Greens-functions}, it is instructive to consider the simpler case of a free spin-1 theory for which the analytic solutions are known for the Green's functions. The single-particle relativistic Green's function in this case is given in Eq. (\ref{eq:G-free-Minkowski}).  Once analytically continued to Euclidean spacetime, this Green's function can be evaluated in terms of modified Bessel functions,   
\begin{eqnarray}
G(\bm{x},\tau;\bm{x}',\tau') &=& \left(\frac{M}{2\pi}\right)^2 \left [ -\frac{d}{d\tau}+M \sigma_3-(\sigma_3+i\sigma_2)\frac{\bm{\nabla}^2}{2M}+\frac{i\sigma_2}{M} (\bm{S} \cdot \bm{\nabla})^2 \right ]
\frac{K_1(Mr)}{Mr}.
\label{eq:G-free-Euclidean}
\end{eqnarray}
Here $r \equiv \sqrt{(\tau-\tau')^2+(\bm{x}-\bm{x}')^2}$ is the Euclidean separation between $x$ and $x'$, and time and space derivatives are taken with respect to unprimed coordinates. According to Eq. (\ref{eq:SR-GR-to-NR-GR}), to make the connection to the NR theory, the following transformation must be performed on $G(\bm{x},t;\bm{x}',t')$
\begin{eqnarray}
G^{(\pm)}_{M_S,M_S'}(\bm{x},\tau;\bm{x}',\tau') \equiv \mathcal{P}^{(\pm)}\otimes \mathcal{T}_{(M_S)}\left[\mathcal{U}^{(1)}(\widehat{\bm{p}})G(\bm{x},\tau;\bm{x}',\tau')\mathcal{U}^{(1)}(\widehat{\bm{p}}')^{-1} \right] \mathcal{P}^{(\pm)}\otimes \mathcal{T}^{T}_{(M_S')},
\label{eq:G-NI-transformed-I}
\end{eqnarray}
with $\mathcal{P}^{(\pm)}$ and $\mathcal{T}_{(M_S)}$ defined after Eq. (\ref{eq:SR-GR-to-NR-GR}), and where $\widehat{\bm{p}_i}(')=-i\frac{d}{d\bm{x}_i(')}$. In the case of a free theory, the exact FWC transformation that brings the relativistic Hamiltonian to a diagonal form to all orders in $1/M$ is known, see Ref. \cite{PhysRev.95.1323}. However for the present purpose, it suffices to consider only the leading order transformation, $\mathcal{U}^{(1)}(\widehat{\bm{p}}) \equiv e^{-i\mathcal{S}^{(1)}(\widehat{\bm{p}})}$, with $\mathcal{S}^{(1)}$ being defined in Eq. (\ref{eq:S-1-def}). Since the free Green's function is translationally invariant, it follows that ${\mathcal{U}^{(1)}}(\widehat{\bm{p}}')G(\bm{x},\tau;\bm{x}',\tau')={\mathcal{U}^{(1)}}(\widehat{\bm{p}})G(\bm{x},\tau;\bm{x}',\tau')$, and as a result Eq. (\ref{eq:G-NI-transformed-I}) can be evaluated to
\begin{eqnarray}
G^{(\pm)}_{M_S,M_S'}(\bm{x},\tau;\bm{x}',\tau')=\mathcal{P}^{(\pm)}\otimes \mathcal{T}_{(M_S)} \left(\frac{M}{2\pi}\right)^2 \left [ -\frac{d}{d\tau} \pm M  \mp \frac{\bm{\nabla}^2}{2M} \right ]
\frac{K_1(Mr)}{Mr} \mathcal{P}^{(\pm)}\otimes \mathcal{T}^{T}_{(M_S')}.
\label{eq:G-NI-transformed-II}
\end{eqnarray}
It can now be observed that
\begin{eqnarray}
\lim_{M \rightarrow \infty,v \rightarrow 0} \left(\frac{M}{2\pi}\right)^2 \left [ -\frac{d}{d\tau} \pm M  \mp \frac{\bm{\nabla}^2}{2M} \right ]
\frac{K_1(Mr)}{Mr} = \pm \left(\frac{M}{\pm 2\pi (\tau - \tau')} \right)^{3/2} e^{\pm M (\tau-\tau') \pm \frac{M(\bm{x}-\bm{x}')^2}{2(\tau-\tau')}},
\label{eq:G-NI-transformed-limit-I}
\end{eqnarray}
when $\pm(\tau-\tau')>0$, and 
\begin{eqnarray}
\lim_{M \rightarrow \infty,v \rightarrow 0} \left(\frac{M}{2\pi}\right)^2 \left [ -\frac{d}{d\tau} \pm M  \mp \frac{\bm{\nabla}^2}{2M} \right ]
\frac{K_1(Mr)}{Mr} = 0,
\label{eq:G-NI-transformed-limit-II}
\end{eqnarray}
when $\pm(\tau-\tau')<0$. Recall that $v={|\bm{x}-\bm{x}'|}/{|\tau-\tau'|}$. These are indeed the NR Green's functions corresponding to positive- and negative-energy modes of the noninteracting theory, see Sec. \ref{sec:NR-Greens-functions}. This therefore confirms the following relation between relativistic and NR Green's functions,
\begin{eqnarray}
\mathcal{G}^{(\pm)}_{M_S,M_S'}(\bm{x},\tau;\bm{x}',\tau')=\lim_{M \rightarrow \infty,v \rightarrow 0} G^{(\pm)}_{M_S,M_S'}(\bm{x},\tau;\bm{x}',\tau').
\label{eq:G-NI-transformed-limit-III}
\end{eqnarray}

When the FV version of this relation is considered with PBCs, the appropriate NR limit must be taken such that the modes that effectively originate from images of the source propagate only with  NR velocities. Since for systems with localized wavefunctions the FV Green's function is simply a sum over all the images of the infinite-volume Green's function, one has
\begin{eqnarray}
\mathcal{G}^{(\pm),FV}_{M_S,M_S'}(\bm{x},\tau;\bm{x}',\tau')= \sum_{\bm{n},\nu}\left[ \lim_{M \rightarrow \infty,v_{\bm{n},\nu} \rightarrow 0} G^{(\pm)}_{M_S,M_S'}(\bm{x}+\bm{n}L,\tau+\nu T;\bm{x}',\tau') \right],
\label{eq:G-NI-transformed-limit-FV}
\end{eqnarray}
where $v_{\bm{n},\nu}={|\bm{x}+\bm{n}L-\bm{x}'|}/{|\tau+\nu T-\tau'|}$, $\nu$ is an integer as before, $\bm{n}$ denotes a three-vector with integer components, and $T$ and $L$ denote the finite temporal and spatial extents of the volume, respectively. Identifying regions of $\bm{x}-\bm{x}'$ and $\tau-\tau'$ in which the NR limit, corresponding to $v_{\bm{n},\nu} \rightarrow 0$, holds requires a straightforward inspection of the relation in Eq. (\ref{eq:G-NI-transformed-limit-FV}). We first note that for all values of $\nu$, $|\tau+\nu T-\tau'|$ will remain large in the region where $|\tau-\tau'|$ asymptotes the value $\frac{T}{2}$. In this region, only nonnegative (nonpositive) values of $\nu$ make nonvanishing contributions to the sum over images in Eq. (\ref{eq:G-NI-transformed-limit-FV}) with $(\tau-\tau')>0$ ($(\tau-\tau')<0$). These correspond to the positive- (negative-)energy Green's functions of the NR theory in a finite volume, see discussions after Eq. (\ref{eq:NR-GF-upper-lower-FV}). The sum over spatial images may seem to create more subtleties when the NR limit is considered. However, once $|\bm{x}_i-\bm{x}_i'|$ asymptotes either the value $0$ or $L$, one remains in the NR region as long as the spatial extent of the volume is large compared with the inverse mass of the state. The reason is that although $|\bm{x}_i-\bm{x}_i'+\bm{n}_i L|$ in not necessarily small for all values of $\bm{n}_i$, the corresponding contributions to the Green's function are exponentially suppressed by at least $e^{-ML}$ in the large volume limit. To summarize, only in the above-mentioned regions for $\bm{x}-\bm{x}'$ and $\tau-\tau'$ may one attempt to match the transformed relativistic Green's functions to the Green's functions of the NR theory. Once external fields are introduced, the identification of the NR region is more involved, nonetheless as long as the applied field is weak, and that the infinite-volume Green's functions in the presence of external fields are still localized, the deviations from the regions discussed above will not be significant.

\bibliography{bibi}
\end{document}